\documentclass[12pt]{JHEP3}
\usepackage{mathrsfs}
\usepackage{amsmath,amssymb}
\usepackage{epsfig}
\usepackage{relsize}
\usepackage{graphicx}

\input epsf

\def\x'{\mathaccent 19 x}
\def\y'{\mathaccent 19 y}
\def\n'{\mathaccent 19 n}
\def\u'{\mathaccent 19 u}

\def\et'{\mathaccent 19 \eta}
\def\th'{\mathaccent 19 \theta}
\def\lam'{\mathaccent 19 \lambda}
\def\varet'{\mathaccent 19 \vartheta}
\def\rh'{\mathaccent 19 \rho}
\def\ph'{\mathaccent 19 \phi}
\def\xb'{\mathaccent 19 {\bar{x}}}

\def\bQ{\overline{Q}}
\def\bC{C^\dagger}

\def\hP{\hat{P}}

\def\l{{\lambda}}

\def\sl(2){\alg{sl}(2)}

\def\be{\begin{equation}}
\def\ee{\end{equation}}

\newcommand{\bea}{\begin{eqnarray}}
\newcommand{\eea}{\end{eqnarray}}

\def\a {\alpha}
\def\b {\beta}

\def\pa {\partial}

\def\g {\gamma}
\def\om {\omega}

\def\la{\label}
\def\e{\epsilon}
\def\ov{\over}

\def\S{\Sigma}

\def\cA{{\cal A}}
\def\dM{\dot{M}}
\def\dN{\dot{N}}
\def\dK{\dot{K}}
\def\dL{\dot{L}}
\def\hS{\widehat{S}}
\def\hP{{\bf P}}
\def\hJ{{\bf J}}
\def\bI{{\bf I}}
\def\bJ{{\bf J}}
\def\bJ{{\bf J}}
\def\bS{{\bf S}}
\def\bR{{\bf R}}
\def\bL{{\bf L}}
\def\bQ{{\bf Q}}
\def\bC{{\bf C}}

\def\bH{{\bf H}}
\def\ba{{\bf a}}
\def\bb{{\bf b}}
\def\tzeta{\widetilde{\zeta}}
\def\ttz{\widetilde{\widetilde{\zeta}}}
\def\ttzeta{\widetilde{\widetilde{\zeta}}}
\def\vp{\varphi}

\newcommand{\alg}[1]{\mathfrak{#1}}
\newcommand{\su}{\alg{su}}

\newcommand{\psu}{\alg{psu}}

\newcommand{\AdS}{{\rm  AdS}_5\times {\rm S}^5}
\newcommand{\ads}{{\rm  AdS}_5\times {\rm S}^5}

\newcommand{\bem}{\left (\begin{matrix}}
\newcommand{\eem}{\end{matrix} \right )}

\def\S{{\cal S}}
\author{Gleb Arutyunov$^a$\footnote{Email: G.Arutyunov@phys.uu.nl, frolovs@maths.tcd.ie, marzam@aei.mpg.de} {}\footnote{Correspondent fellow at Steklov
Mathematical Institute, Moscow.}\, , Sergey Frolov$^{b\, \dagger}$
and Marija Zamaklar$^{c}$ \\ $^{a}$ {\it Institute for Theoretical
Physics and Spinoza Institute,\\ ~~Utrecht University, 3508 TD
Utrecht, The Netherlands} \\ $^b$ {\it School of Mathematics,
Trinity College, Dublin 2, Ireland} \\  $^{c}$ {\it
Max-Planck-Institut f\"ur Gravitationsphysik,
Albert-Einstein-Institut\\ ~~Am M\"uhlenberg 1, D-14476 Potsdam,
Germany}}

\abstract{ We discuss the Zamolodchikov-Faddeev algebra for the
superstring sigma-model on $\AdS$. We find the \emph{canonical} $\su(2|2)^2$
invariant S-matrix satisfying  the standard Yang-Baxter
and crossing symmetry equations. Its near-plane-wave expansion
matches exactly the leading order term recently
obtained by the direct perturbative
computation. We also show that the S-matrix obtained by Beisert in
the gauge theory framework does not satisfy the standard
Yang-Baxter equation, and, as a consequence,  the
corresponding ZF algebra is twisted. The S-matrices in gauge and
string theories however are physically equivalent and related by a
non-local transformation of the basis states which is explicitly
constructed. }

\title{The Zamolodchikov-Faddeev Algebra \\for   $\AdS$
Superstring}

\preprint{\smaller{\smaller{\smaller{AEI-2006-099}}}\\[-.5ex]
          \smaller{\smaller{\smaller{ITP-UU-06-58}}}\\[-.5ex]
          \smaller{\smaller{\smaller{SPIN-06-48}}}\\[-.5ex]
          \smaller{\smaller{\smaller{TCDMATH 06-18}}}}

\begin{document}

\renewcommand{\thefootnote}{\arabic{footnote}}
\setcounter{footnote}{0}
\section{Introduction and summary}

Unravelling the integrable properties of the $\AdS$ string sigma
model \cite{MeT} and of the dual gauge theory
\cite{Bena:2003wd,Minahan:2002ve} allowed for the remarkable progress in
understanding the spectra and interrelation of these theories.
Most importantly, determination of the spectrum in the large
volume (charge) limit was shown to rely on the knowledge of the
S-matrix describing the scattering of world-sheet excitations, or,
alternatively, excitations of a certain spin chain in the dual
gauge theory \cite{AFS,S,B}.

\medskip

It turned out that the form of the S-matrix is severely restricted
by the requirement of invariance under the global symmetries of
the model. It was argued in \cite{B} and explicitly derived in
\cite{AFPZ}, that the relevant symmetry algebra of the gauge-fixed
off-shell string sigma model in the infinite volume limit is the
centrally extended $\psu(2|2)\oplus \psu(2|2)$ superalgebra.
Requiring the S-matrix to be invariant under this symmetry algebra
fixes its form essentially uniquely up to an overall phase
\cite{B,Bn}. Constraints on this phase were established in
\cite{Janik} by demanding the S-matrix to satisfy the crossing
relation, the common property of  S-matrices in relativistic field
theories. A physically relevant solution of the crossing relation,
which successfully reproduces the known string theory data was
conjectured in \cite{BHL}, building on the previous work
\cite{Beisert:2005cw}-\cite{FrK}. Remarkably, the same
solution for the crossing-symmetric phase was shown to arise from
perturbative gauge theory \cite{BES}. A number of further
important verifications has been made
\cite{Bern}-\cite{Maldacena:2006rv}, \footnote{It is still unclear
if the proposal of \cite{BES} correctly describes bound states
present in string and gauge theories at large values of the 't
Hooft coupling constant $\lambda$ \cite{HM}. Then the resulting
Bethe ansatz is supposed to be asymptotic and might not capture
finite size corrections
 \cite{AJ,SZZ,Hub,magnon}.} which supports the picture of the
interpolating crossing-symmetric S-matrix from weak (gauge theory)
to strong (string theory) coupling.

\medskip

Bearing in mind all these highly exciting developments, a further
question to ask is what is a proper characterization of elementary
world-sheet excitations whose scattering is governed by the
crossing-symmetric S-matrix? In other words, we would like to
understand the structure of the Hilbert space of string theory
arising in the infinite volume (charge) limit.

\medskip

Although the gauge-fixed string theory we are interested in does
not possess relativistic invariance on the world-sheet, it is
useful to invoke an analogy with  two-dimensional massive
integrable models. The basic feature of these models is the
existence of infinite number of commuting conservation laws which
lead to the {\it factorized scattering} preserving both a number
of particles and the set of their on-shell momenta \cite{Zam}. The
space of local operators is specified by the two-body S-matrix, and
the Hilbert space of asymptotic states forms a representation of
the Zamolodchikov-Faddeev (ZF) algebra \cite{Zam,Fad}. In
particular, the asymptotic states diagonalize the local
conservation laws. Thus, the ZF algebra provides an indispensable
framework for treating massive integrable models. It also allows
\cite{Lukyanov:1993pn}, through the bosonization procedure, the
computation of the corresponding form-factors \cite{Smirnov}.

\medskip

In this paper we discuss the ZF algebra for the superstring
sigma-model on $\AdS$. In the string theory context the ZF algebra
represents a natural generalisation of the free field oscillators
which describe elementary excitations in the plane-wave limit
\cite{Berenstein:2002jq} of the world-sheet theory. As
interactions are switched on integrability ensures that the Fock
structure of the Hilbert space is preserved. The effect of
interactions is then taken into account by deforming the algebra
of creation and annihilation operators with the help of a
non-trivial scattering matrix. Thus, in momentum space the ZF
algebra is generated by the operators $A(p)$ and $A^{\dagger}(p)$
satisfying the following relations
\begin{equation}\
A_1A_2 = {\cal S}_{12}A_2A_1\,,\quad A^\dagger_1A^\dagger_2=
A^\dagger_2A^\dagger_1{\cal S}_{12}\,,\quad A_1A^\dagger_2=
A^\dagger_2 {\cal S}_{21}A_1 + \delta_{12}\,,\\\nonumber
\end{equation}
where ${\cal S}_{12}$ is the two-body S-matrix of the model and
$\delta_{12}$ is the delta-function depending on the difference of
momenta of scattering particles.

\medskip

We thus see that the construction of the ZF algebra relies on the
knowledge of the two-body S-matrix. Moreover, the consistency of
the ZF algebra relations implies that the S-matrix should obey the
Yang-Baxter (YB) equation. In this paper we show that such an S-matrix
consistent with the symmetries of the gauged-fixed string model
does exist and we find its explicit form. This \emph{canonical} $\su(2|2)^2$ invariant S-matrix exhibits
the following properties
\begin{itemize}
\item it obeys the standard Yang-Baxter equation
\begin{equation}
\nonumber
 {\cal S}_{23}{\cal S}_{13}{\cal S}_{12}={\cal S}_{12}{\cal S}_{13}{\cal S}_{23}\,
\end{equation}

\item it obeys the unitarity condition
\begin{equation}
\nonumber {\cal S}_{12}(p_1,p_2){\cal S}_{21}(p_2,p_1)={\mathbb
I}\,
\end{equation}

\item it obeys physical unitarity condition which is sometimes referred to
as ``hermitian analyticity''
\begin{equation}
\nonumber {\cal S}_{21}^{\dagger}(p_2,p_1)= {\cal
S}_{12}(p_1,p_2)\,
\end{equation}

\item it obeys the requirement of crossing symmetry
\bea \nonumber \mathscr{C}_1^{-1}{\cal S}_{12}^{t_1}(p_1,p_2)
\mathscr{C}_1
 {\cal S}_{12}(-p_1,p_2)={\mathbb I}\, , \eea
where $\mathscr{C}$ is the charge conjugation matrix.

\end{itemize}
We will show that all these requirements on the S-matrix naturally
follow from the consistency conditions of the ZF algebra. In
particular, the requirement of crossing symmetry reflects the
compatibility of the factorized scattering with the ${\mathbb
Z}_4$-graded structure of the superalgebra $\psu(2|2)$ and it
results into a non-trivial equation on the overall phase of the
S-matrix which is the same equation as derived in \cite{Janik}.

\medskip

The S-matrix we consider depends on the string tension. We then
show that expanding ${\cal S}$ in the large tension limit we
recover the near plane-wave S-matrix which was recently obtained
from the gauge-fixed string sigma-model in \cite{KMRZ}. Moreover,
we show that in accordance with \cite{AF06}, the overall phase of
${\cal S}$ is capable to incorporate the various gauge choices
compatible with the symmetries of the string model.

\medskip

We find it pretty remarkable that the sigma-model describing
$\AdS$ superstring admits the S-matrix which shares most of the
usual properties of S-matrices arising in relativistic
two-dimensional massive integrable models. The point is that in
the dual gauge theory the spin chain which describes the local
composite gauge-invariant operators is dynamic because the
scattering process requires fluctuations of the length
\cite{Beisert:2003ys,B}. Consequently, one would expect that the
corresponding S-matrix would not obey the usual Yang-Baxter
equation but rather a new dynamic, or ``twisted" Yang-Baxter
equation. Indeed, some examples of integrable systems are known
which are naturally described in terms of a dynamic S-matrix
satisfying a twisted version of the Yang-Baxter  equation, see e.g.
\cite{GNF}.

\medskip

We have to bear in mind, however, that the form of the S-matrix
and, consequently, its properties, do depend on the choice of the
scattering basis. Performing ``gauge" transformations on the
two-particle scattering basis we transform  the S-matrix as well.
Note that in general we should allow for transformations of the
two-particle scattering basis, which from the point of view of
one-particle states, look mutually non-local. These
transformations might significantly modify the properties of the
S-matrix without changing the actual physical content. In
particular, there could exist a preferable basis in the space of
two-particle states in which the S-matrix exhibits the standard
properties. In fact, it is precisely the basis provided by string
theory in which we find ${\cal S}(p_1,p_2)$ with the nice
properties listed above.

\medskip

Our construction of the $\su(2|2)$-invariant S-matrix is the
conventional field-theoretic one. We start with the discussion of
the usual ZF algebra and show that invariance of the ZF relations
acting on the vacuum state under the action of the $\su(2|2)$
symmetry algebra\footnote{Since the symmetry algebra of the
gauge-fixed string Hamiltonian is a sum of two copies of
$\psu(2|2)$ it is sufficed to discuss only one. Whenever it is
impossible to treat the ``chiral sectors" separately we will make
a comment.} translates into a certain invariance condition for the
S-matrix which is essentially the same as in \cite{Bn}. It is
worthwile to note that due to the non-linear nature of the
dynamical supersymmetry generators \cite{AFPZ} these generators
have a non-trivial braiding relations with the ZF operators. The
braiding factors involve the operator of the world-sheet momentum
$\hP=\int {\rm d}p~p\,A_i^{\dagger}(p)A^i(p)$ which also plays the
role of the central charge for the symmetry algebra in question.
Such a realization of non-linear symmetries is well known in the
theory of integrable systems \cite{Bernard:1990ys} and it has been
recently emphasized in the context of the $\AdS$ string
sigma-model in \cite{KMRZ}.

\medskip

In a generic two-particle basis the S-matrix, $S_{12}\equiv
S(p_1,p_2)$, found from the $\su(2|2)$-invariance condition satisfies
the twisted Yang-Baxter equation whose form depends of the twist
matrix $F$:
\begin{equation}
\nonumber
  F_{23}(p_1) S_{23}F_{23}^{-1}(p_1)S_{13}F_{12}(p_3)S_{12}F^{-1}_{12}(p_3)
=S_{12}F_{13}( p_2)S_{13}F_{13}^{-1}(p_2)S_{23}\, .
\end{equation}
We find that this condition can be naturally interpreted as the
consistency condition for the twisted ZF algebra based on the
operator S-matrix $\hS_{12}$:
\begin{equation}
\nonumber \hS_{12} =F_{12}(\hP) S_{12} F^{-1}_{12}(\hP) \, .\quad
\end{equation}
We show, however,  that there is a very special basis such that
bosonic and fermionic ZF operators are transformed by
supersymmetry generators in the same fashion, i.e. it is the basis which preserves a symmetry between two $\su(2)$-factors of
$\su(2|2)$.
We refer to this basis as to the ``string'' one
because in this basis the action of the symmetry generators is the
same as implied by the gauge-fixed string \mbox{sigma-model
\cite{AFPZ}.} We find that in the string basis the S-matrix
coincides with ${\cal S}(p_1,p_2)$, and, as was discussed above,
satisfies the usual YB equation. We then show that the
transformation of the ZF basis to a generic one involves the
momentum operator $\hP$, and as a result twists the ZF algebra and
the corresponding YB equation. Finally, we also find a
relation of ${\cal S}(p_1,p_2)$ to the one obtained in \cite{B,Bn}.

\medskip

Thus, we show that $\AdS$ string sigma-model can be naturally
embedded in the general framework of massive integrable systems.
Of course, the S-matrix we find has also properties which are
different from those in the relativistic integrable QFTs. In
particular, it does not depend on the difference of rapidities of
scattering particles which reflects a non-relativistic nature of
our model. Then the crossing symmetry relation implies that ${\cal
S}(p_1,p_2)$ cannot be a meromorphic function of the particle
momenta \cite{BHL,Maldacena:2006rv}.

\medskip

Concluding this section we note several open questions. First, it
would be interesting  to work out  the details of the nested Bethe
ansatz based on ${\cal S}(p_1,p_2)$.
%Second, one would like to
% understand better the implementations of the Hopf algebra
%symmetries \cite{Janik,Gomez:2006va} within  our approach.
Second, it is desirable to construct realization of the symmetry
generators of $\su(2|2)$ via the oscillators of the ZF algebra.
This would allow one to put the ZF algebra in the corner of the
axiomatic construction. Since the S-matrix we found obeys most of
the standard properties of the S-matrices in massive relativistic
integrable systems it is tempting to use the ideas of the
thermodynamic Bethe ansatz \cite{za} to describe the finite size effects.
Finally, it would be interesting to understand the relation of our
approach to those of \cite{MP}-\cite{Gromov:2006dh}.

\section{S-matrix and symmetries}
\subsection{S-matrix}
In scattering theory, the S-matrix is a unitary operator, which we denote $\bS$, mapping free particle $out$-states to free particle $in$-states in the Heisenberg picture. In Dirac notation, we define $|0\rangle$ as the  vacuum quantum state. If
$A^\dagger_i(p)$ is a
 creation operator, its hermitian conjugate is the vacuum annihilation operator:
\bea\nonumber
A^i(p)|0\rangle =0\,.
\eea
To describe the scattering  process we introduce the $in$-basis and the $out$-basis as follows
\bea\nonumber
 &&|p_1,p_2, \cdots , p_n \rangle^{(in)}_{i_1,...,i_n} =A^\dagger_{i_1}(p_1)\cdots  A^\dagger_{i_n}(p_n)|0 \rangle  \,,\quad p_1>p_2>\cdots >p_n\,,\\\nonumber
 &&|p_1,p_2, \cdots , p_n \rangle^{(out)}_{i_1,...,i_n} =A^\dagger_{i_n}(p_n)\cdots  A^\dagger_{i_1}(p_1)|0 \rangle  \,,\quad p_1>p_2>\cdots >p_n\,.
\eea Then in the scattering process the $in$ state goes to the
$out$ state \bea\nonumber
|p_1,...,p_n\rangle^{(in)}_{i_1,...,i_n}\rightarrow
|p_1,...,p_n\rangle^{(out)}_{i_1,...,i_n}\,. \eea Both states
belong to the same Hilbert space and we can expand initial states
on a basis of final states. In particular, the two-particle $in$
and $out$ states are related as follows: \bea\nonumber
|p_1,p_2\rangle^{(in)}_{i,j} = \bS\cdot
|p_1,p_2\rangle^{(out)}_{i,j}=S^{kl}_{ij}(p_1,p_2)
|p_1,p_2\rangle^{(out)}_{k,l}\,, \eea or by using the explicit
basis we can write \bea\la{f1} A^\dagger_{i}(p_1)
A^\dagger_{j}(p_2)|0 \rangle =\bS\cdot A^\dagger_{j}(p_2)
A^\dagger_{i}(p_1)|0 \rangle= S^{kl}_{ij}(p_1,p_2)
A^\dagger_{l}(p_2)  A^\dagger_{k}(p_1)|0 \rangle\,. \eea The usual
ZF algebra then follows just by dropping $|0 \rangle$ on both
sides of the equation \bea\la{ZFu} A^\dagger_{i}(p_1)
A^\dagger_{j}(p_2) =  A^\dagger_{l}(p_2)
A^\dagger_{k}(p_1)S^{kl}_{ij}(p_1,p_2)\,. \eea It is clear from
this form of the algebra that in the absence of interaction it
should be equal to the graded unit matrix, that is to the diagonal
matrix with $\pm 1$ depending on the statistics of the
corresponding creation operator. In addition, in most of the
two-dimensional integrable models the S-matrix is equal to the
minus (usual!) permutation matrix for $p_1=p_2$, because a
two-particle state describes two solitons moving with momenta
$p_1$ and $p_2$, and there is no two-soliton state with equal
momenta.

Then, the consistency condition for the ZF algebra  leads to the YB equation
\bea\la{YBu}
S_{23}S_{13}S_{12}
=S_{12}S_{13}S_{23}\,,  \eea
where we have used the standard tensor notations, see Appendix \ref{notations}.

It is clear, however, that it is not the most general form of the ZF algebra, for
example a tensor operator $U$ leaving the vacuum invariant  could
appear on the rhs of eq.(\ref{ZFu}) \bea\la{ZFg}
A^\dagger_{i}(p_1)  A^\dagger_{j}(p_2) = S^{kl}_{ab}(p_1,p_2)
A^\dagger_{n}(p_2)  A^\dagger_{m}(p_1)U_{ij,kl}^{ab,mn}\,,\quad
U_{ij,kl}^{ab,mn}|0 \rangle
=\delta_i^a\delta_j^b\delta_k^m\delta_l^n |0 \rangle\,.~~~\eea
The
appearance of the operator $U$ modifies the consistency condition
and the YB equation leading to the ``twisted'' ZF algebra and YB equation. We will see  that this kind of modification occurs
for strings in $\AdS$ if one uses in the ZF algebra (\ref{ZFg}) the $\su(2|2)^2$  spin chain  S-matrix found in \cite{B}.  On the other hand,
the operator $U$ can also appear because of an inconvenient choice of the basis of the creation/annihilation operators in the ZF algebra. In that case, one can find a proper basis and untwist the twisted ZF algebra.
We will see  that this is what happens for strings in $\AdS$, and find the right and very natural from string theory point of view basis of the ZF algebra, and the \emph{canonical} $\su(2|2)^2$ invariant S-matrix satisfying the usual YB equation.

%%%%%%%%%%%%%%%%%%%%%%%%%%%%%
\subsection{Symmetries}
%%%%%%%%%%%%%%%%%%%%%%%%%%%%%%
Let us assume that the Hamiltonian commutes with a set of symmetry
algebra generators  $\hJ^\ba$ which are operators acting in the
Hilbert space of states. Then the Hilbert space of states carries
a linear representation of the symmetry algebra. This implies in
particular that for generators preserving the number of particles
\bea \la{strcon} &&\hJ^\ba\cdot|0 \rangle =0\,,\\\nonumber
&&\hJ^\ba\cdot A^\dagger_{i}(p)|0 \rangle =
J^{\ba}{}^j_{i}(p)A^\dagger_{j}(p)|0 \rangle\,,\\\nonumber
&&\hJ^\ba\cdot A^\dagger_{i}(p_1)  A^\dagger_{j}(p_2)|0 \rangle =
J^\ba{}^{kl}_{ij}(p_1,p_2)A^\dagger_{k}(p_1)  A^\dagger_{l}(p_2)|0
\rangle\,. \eea Let us stress that the two-particle state
$A^\dagger_{i}(p_1)  A^\dagger_{j}(p_2)|0 \rangle$ cannot in
general be identified with $A^\dagger_{i}(p_1)|0 \rangle\otimes
A^\dagger_{j}(p_2)|0 \rangle$ equipped with the standard action of
the symmetry generators because in that case the representation
constants would satisfy \bea\nonumber J^\ba{}^{kl}_{ij}(p_1,p_2) =
J^\ba{}^{k}_{i}(p_1)\delta_j^l+\delta_i^k J^\ba{}^{l}_{j}(p_2)\,.
\eea The invariance condition for the S-matrix can be derived
multiplying (\ref{f1}) by $\hJ^\ba$ \bea \hJ^\ba\cdot
A^\dagger_{i}(p_1)  A^\dagger_{j}(p_2)|0 \rangle =
S^{kl}_{ij}(p_1,p_2)\,\hJ^\ba\cdot  A^\dagger_{l}(p_2)
A^\dagger_{k}(p_1)|0 \rangle\,. \eea Computing the lhs and rhs of
this equality, we see that the S-matrix must satisfy the following
invariance condition \bea\la{incon}
S^{mn}_{kl}(p_1,p_2)J^\ba{}^{kl}_{ij}(p_1,p_2)=
J^\ba{}^{nm}_{lk}(p_2,p_1)S^{kl}_{ij}(p_1,p_2)\,. \eea If we
combine the symmetry generator structure constants in a matrix
\bea J^\ba_{12}(p_1,p_2)\equiv J^\ba{}^{kl}_{ij}(p_1,p_2) E^i_k
\otimes E^j_l\,, \eea where $E_k^i$ are the standard matrix
unities, c.f. appendix 8.1, then the invariance condition can be
written as \bea\la{incon2} S_{12}(p_1,p_2)J^\ba_{12}(p_1,p_2) =
J^\ba_{21}(p_2,p_1)S_{12}(p_1,p_2)\,. \eea
 Note that the condition reduces to the familiar one
\bea
[J^a\otimes \mathbb{I}  + \mathbb{I}\otimes J^a,S_{12}]=0\,,
\eea
only if $J^a(p) = J^\ba{}^{k}_{i}(p)E^i_k$ is independent of $p$.

\medskip

The form of $J^\ba_{12}(p_1,p_2)$ depends on the symmetry algebra of a particular model. If the symmetry algebra is equipped with the structure of a bi-algebra, then the two-particle representation is given by the bi-algebra coproduct. If the symmetry algebra has a center then multi-particle representations are characterized by the values of the central elements, and a  two-particle state may be considered as the tensor product of two one-particle states with in general different values of the central elements. Then, if a symmetry algebra is a Lie algebra, a two-particle representation is given by the standard coproduct, and the structure constants have the following form
\bea\la{twopart}
J^\ba{}^{kl}_{ij}(p_1,p_2) = J^\ba{}^{k}_{i}(p_1;c_1)\delta_j^l+(-1)^{\epsilon(i)\epsilon(\ba )}\delta_i^k J^\ba{}^{l}_{j}(p_2;c_2)\,,
\eea
where $c_i$ denote sets of central charges which characterize the ``one-particle'' representations, and $\epsilon(i)=0$ if $A^\dagger_i$ is boson, and $\epsilon(i)=1$ if $A^\dagger_i$ is fermion, and the same holds for $\epsilon(\ba )$. The structure constants $J^\ba{}^{k}_{i}(p)$ in (\ref{strcon}) of one-particle states correspond to definite values of the central charges, say,  $c_i=0$. Since a two-particle state represents two asymptotic noninteracting particles, the central charges $c_i$ may depend only on the coupling constants and the values of the momenta $p_i$ of the particles.

Two-particle structure constants of the form (\ref{twopart}) which depend only on momenta of the particles can be derived by assuming the following commutation relations of the symmetry algebra generators with the creation operators
\bea\la{JA}
\hJ^\ba A^\dagger_i(p) =J^\bb{}^{k}_{m}(p)A^\dagger_k(p) \Theta^{\ba m}_{\bb i}(p;\hP)+(-1)^{\epsilon(i)\epsilon(\ba )} A^\dagger_m(p)\widetilde{\Theta}^{\ba m}_{\bb i}(p;\hP)\hJ^\bb \,.
\eea
Here $\hP$ is the conserved world-sheet momentum of the model obeying the following simple commutation relation
\bea\nonumber
\hP A^\dagger_i(p)=A^\dagger_i(p) \left( \hP +p  \right)\,,
\eea
 and the braiding operators $\Theta^{\ba m}_{\bb i}(p;\hP)$ and $\widetilde{\Theta}^{\ba m}_{\bb i}(p;\hP)$ should satisfy the following conditions
\bea\nonumber
&&\Theta^{\ba m}_{\bb i}(\hP)|0 \rangle = \delta^\ba_\bb \delta^m_i|0 \rangle\,,\\\nonumber
&&J^\bb {}^{k}_{m}(p_1)A^\dagger_k(p_1)\Theta^{\ba m}_{\bb i}(p_1;\hP)A^\dagger_j(p_2)|0 \rangle = J^\ba{}^{k}_{i}(p_1;c_1)A^\dagger_k(p_1)A^\dagger_j(p_2)|0 \rangle\,,\\\nonumber
&&A^\dagger_m(p_1)\widetilde{\Theta}^{\ba m}_{\bb i}(p_1;\hP)\hJ^\bb A^\dagger_j(p_2)|0 \rangle = J^\ba{}^{k}_{j}(p_2;c_2)A^\dagger_i(p_1)A^\dagger_k(p_2)|0 \rangle\,.
\eea
The last two conditions are equivalent to
\bea\nonumber
J^\bb{}^{k}_{m}(p_1)\Theta^{\ba m}_{\bb i}(p_1;p_2) = J^\ba{}^{k}_{i}(p_1;c_1)\,,\quad
\widetilde{\Theta}^{\ba m}_{\bb i}(p_1;p_2)J^\bb{}^{k}_{j}(p_2)
= \delta^m_iJ^\ba{}^{k}_{j}(p_2;c_2)\,.~~~
\eea
It is clear that the braiding operators should also satisfy the consistency conditions that follow from the fact that the symmetry operators form an algebra. In what follows we restrict our attention to the following choice of (\ref{JA}) which is compatible with the ZF algebra we will discuss in next sections
\bea\la{JAs}
\hJ^\ba A^\dagger_i(p) =J^\bb{}^{k}_{m}(p)A^\dagger_k(p) \Theta^{\ba m}_{\bb i}(\hP)+(-1)^{\epsilon(i)\epsilon(\ba )} A^\dagger_i(p)\hJ^\ba \,,
\eea
where the only braiding operator $\Theta^{\ba m}_{\bb i}(\hP)$ depends only on the world-sheet momentum operator $\hP$, and is equal to $\delta^\ba_\bb\delta^m_i$ if $\hJ^\ba$ is bosonic.

Another simple choice of (\ref{JA}) is as follows
\bea\la{JAb}
\hJ^\ba A^\dagger_i(p) =J^\ba{}^{k}_{m}(p)A^\dagger_k(p)+(-1)^{\epsilon(i)\epsilon(\ba )} A^\dagger_m(p)\widetilde{\Theta}^{\ba m}_{\bb i}(p;\hP)\hJ^\bb \,.
\eea
In general, the corresponding (twisted) ZF algebra would not have the simplest form, and we will not be discussing this choice in detail.

%%%%%%%%%%%%%%%%%%%%%%%%%%%%%%%%%%%%%%%%%%
%%%%%%%%%%%%%%%%%%%%%%%%%%%%%%%%%%%%%%%%%%

\section{The centrally extended $\su(2|2)$ and its representations}

The off-shell symmetry algebra of the light-cone $\AdS$
superstring was recently discussed in detail in \cite{AFPZ} where
it was shown that it consists of two copies of the centrally
extended $\su(2|2)$ algebra, both copies sharing the same central
element which corresponds to the world-sheet light-cone
Hamiltonian. To describe the  ZF algebra we need to recall the
structure of representations of the centrally extended $\su(2|2)$
algebra. In our discussion we  closely follow the one in
\cite{Bn}.

%%%%%%%%%%%%%%%%%%%%%%%%%%%%%%%%%%%%%%

\subsection{Centrally extended $\su(2|2)$ algebra}

%%%%%%%%%%%%%%%%%%%%%%%%%%%%%%%%%%%%%%

The centrally extended $\su(2|2)$ algebra which we will denote $\su(2|2)_{H,C}$,  consists of the rotation generators
$\bL_a{}^b\,,\ \bR_\a{}^\b$ of the $\su(2)\oplus\su(2)$ bosonic subalgebra,  the supersymmetry generators $\bQ_\a{}^a,\,\ \bQ_a^{\dagger}{}^\a$ and
three central elements $\bH$, $\bC$ and $\bC^\dagger$.
\bea \label{su22}
&& \left[\bL_a{}^b,\bJ_c\right]=\delta_c^b \bJ_a - {1\ov 2}\delta_a^b \bJ_c\,,\qquad
\left[\bR_\a{}^\b,\bJ_\g\right]=\delta^\b_\g \bJ_\a - {1\ov 2}\delta^\b_\a \bJ_\g\,,
 \nonumber \\
&& \left[\bL_a{}^b,\bJ^c\right]=-\delta_a^c \bJ^b + {1\ov 2}\delta_a^b \bJ^c\,,\qquad
\left[\bR_\a{}^\b,\bJ^\g\right]=-\delta_\a^\g \bJ^\b + {1\ov 2}\delta_\a^\b \bJ^\g\,, \nonumber \\
&& \{ \bQ_\a{}^a, \bQ_b^\dagger{}^\b\} = \delta_b^a \bR_\a{}^\b + \delta_\a^\b \bL_b{}^a +{1\ov 2}\delta_b^a\delta^\b_\a  \bH\,, \nonumber \\
&& \{ \bQ_\a{}^a, \bQ_\b{}^b\} = \epsilon_{\a\b}\epsilon^{ab}~\bC\,
, ~~~~~~~~ \{ \bQ_a^\dagger{}^\a, \bQ_b^\dagger{}^\b\} =
\epsilon_{ab}\epsilon^{\a\b}~\bC^\dagger \,. \eea Here in the
first two lines we indicate how the indices $c$ and $\gamma$ of
any Lie algebra generator transform under the action of the
bosonic subalgebras generated by $\bL_a{}^b$ and $\bR_\a{}^\b$.
The supersymmetry generators $\bQ_\a{}^a$ and
$\bQ_a^{\dagger}{}^\a$, and the central elements $\bC$ and
$\bC^\dagger$ are hermitian conjugate to each other: $\left(
\bQ_\a{}^a\right)^\dagger=\bQ_a^{\dagger}{}^\a$. The central
element $\bH$ is hermitian and is identified with the world-sheet
light-cone Hamiltonian. It was shown in \cite{AFPZ} that the
central element $\bC$ is expressed through the world-sheet
momentum $\hP$ as follows \bea \label{Cc}
\bC=ig\,(e^{i\hP}-1)e^{2i\xi}\,,\quad g ={\sqrt\l\ov 4\pi}\, .
\eea The phase $\xi$ is an arbitrary function of the central
elements, and reflects the obvious ${\rm U(1)}$ automorphism of
the algebra (\ref{su22}): $\bQ\to e^{i\xi}\bQ\,,\ \bC\to
e^{2i\xi}\bC$. The phase $\xi$  can be fixed to be $0$ by choosing
a proper form of the supersymmetry generators. This is the choice
we use throughout the paper. This simplifies the comparison with
the explicit string theory computation of the S-matrix performed
in \cite{KMRZ}. It is worth mentioning, however, that to have the
interpretation of a multi-particle state as the tensor product of
fundamental representations, it is necessary to consider
fundamental representations with arbitrary phases $\xi$, see the
next section.

%%%%%%%%%%%%%%%%%%%%%%%%%%%%%%%%%%%%%%

\subsection{Fundamental representation}\la{Repr}

%%%%%%%%%%%%%%%%%%%%%%%%%%%%%%%%%%%%%%

The fundamental representation of $\su(2|2)_{H,C}$ is
four-dimensional. We denote the corresponding vector space as
$V(p,\zeta)$, and the basis vectors as $|e_M(p,\zeta)\rangle\equiv
|e_M\rangle$, where the index $M=\{a,\a\}$, and $a=1,2$ are
bosonic indices and $\a=3,4$ are fermionic ones. The parameters
$p$ and $\zeta$ are, for non-unitary representations,  arbitrary
complex numbers which parameterize the values (charges) of the
central elements on this representation: $\bH  |e_M\rangle= H
|e_M\rangle\,,\ \bC   |e_M\rangle =C |e_M\rangle\,,\ \bC^\dagger
|e_M\rangle =\overline{C} |e_M\rangle$. The canonical fundamental
representation of bosonic generators is \bea \nonumber \bL_a{}^b
|e_c\rangle = |e_M\rangle L^{bM}_{ac}= \delta_c^b|e_a\rangle -
{1\ov 2}\delta_a^b|e_c\rangle\,,\qquad \bR_\a{}^\b |e_\g\rangle =
|e_M\rangle R^{\b M}_{\a\g}= \delta_\g^\b|e_\a\rangle -  {1\ov
2}\delta_\a^\b|e_\g\rangle\,, \eea and the most general
fundamental representation of supersymmetry generators is of the
form \bea \nonumber \bQ_\a{}^a |e_b\rangle &=& |e_M\rangle
Q^{aM}_{\a b}=a\, \delta_b^a|e_\a\rangle \,,\qquad \bQ_\a{}^a
|e_\b\rangle = |e_M\rangle Q^{a M}_{\a\b}=
b\,\e_{\a\b}\e^{ab}|e_b\rangle \,,\\ \bQ_a^\dagger{}^\a
|e_\b\rangle &=& |e_M\rangle  \overline{Q}^{\a M}_{a\b}=
d\,\delta_{\b}^{\a}|e_a\rangle \,,\qquad \bQ_a^\dagger{}^\a
|e_b\rangle = |e_M\rangle \overline{Q}^{\a M}_{a b}=c\,
\e_{ab}\e^{\a\b}|e_\b\rangle
 \,. \la{repg}
\eea
Here $a,b,c,d$ are complex numbers parametrizing the representation, and
$Q^{aM}_{\a b}\,, \ Q^{a M}_{\a\b}$ and so on are the representation structure constants, and the corresponding matrices $Q_\a^a$ and so on which appear in the invariance condition (\ref{incon}) for the S-matrix are constructed by using the matrix units defined in Appendix  \ref{notations}:
$Q_\a^a =Q^{aM}_{\a N} e_M^N= Q^{aM}_{\a b} e_M^b + Q^{a M}_{\a\b} e^\b_M$.

The central charges of this representation are expressed through the parameters $a, b, c, d$ as follows
\bea\la{hc1}
H = ad+bc\,,\quad C = ab\,,\quad \overline{C} = cd\,,
\eea
and the consistency condition
\bea
ad-bc=1
\eea
has to be imposed to obtain a representation of the centrally extended $\su(2|2)$.

For generic values of the parameters $a,b,c,d$ the representation is non-unitary. To get a unitary representation one should impose the conditions $d={\bar a}, c =\bar{ b}$.

The parameters $a,b,c,d$ depend on the string sigma model coupling
constant $g$, and the world-sheet momenta $p$. To express this
dependence it is convenient to introduce new parameters
$g,x^+,x^-,\zeta,\eta$ as follows \cite{Bn}\footnote{The parameter
$\zeta$ in \cite{Bn} should be rescaled as $\zeta\to -i\zeta$ to
match our definition.} \bea\la{abcd} a = \sqrt{g}\eta\,,\quad b =
\sqrt{g}{i\zeta\ov \eta}\left( {x^+\ov x^-}-1\right)\,,\quad c =
-\sqrt{g}{\eta\ov \zeta x^+}\,,\quad d = \sqrt{g}{x^+\ov
i\eta}\left( 1-{x^-\ov x^+}\right)\,.~~~ \eea The parameters
$x^\pm$ satisfy the constraint \bea\la{xpm} x^+ +{1\ov x^+}
-x^--{1\ov x^-}={i\ov g}\,, \eea which follows from $ad-bc=1$, and
are related to the momentum $p$ as \bea {x^+\ov x^-} = e^{ip}\,.
\eea The values of the central charges can be found by using
(\ref{hc1}) \bea\la{hc2}
H &=& 1 + {2ig\ov x^+}-  {2ig\ov x^-} =  2ig x^- - 2ig x^+ -1\,,\quad H^2 = 1 + 16 g^2\sin^2{p\ov 2}\, ,\\
C&=&ig\zeta\left( {x^+\ov x^-}-1\right) =ig \zeta\left( e^{ip}-1\right)\,,\quad
 \overline{C}={g\ov i\zeta}\left( {x^-\ov x^+}-1\right) = {g\ov i\zeta}\left( e^{-ip}-1\right)\,.~~~~~
\eea
Comparing the expressions with the formula (\ref{Cc}) derived in string theory \cite{AFPZ}, we see
that the parameter $\zeta$ should be identified with $e^{2i\xi}$, and $p$ with the value of the world-sheet momentum $\hP$ on this representation.

\medskip

The fundamental representation is completely determined by the
parameters $g,x^+,x^-,\zeta$. The parameter $\eta$ just reflects a
freedom in the choice of the basis vectors $|e_M\rangle$. For a
non-unitary fundamental representation it can be set to 1 by
rescaling $|e_M\rangle$ properly. From the string theory point of
view it is not a natural choice of the basis because the
supersymmetry generators act in a different way on bosons and
fermions, and the explicit form of the generators found in
\cite{LCpaper,AFPZ} shows that the action should be similar. The
string theory symmetric choice corresponds to taking $\eta \sim
\sqrt\zeta$.

\subsection{Unitary representation}
In string theory we are interested in unitary representations.
A fundamental unitary representation is singled out by the conditions $c=\bar{b}, d=\bar{a}$, and $p$ is real, and is  given by choosing the parameters
 $\eta$ and $\zeta$ to be
\bea\la{urep}
\zeta
=e^{2i\xi}\,,\quad \eta = \sqrt{i x^- - i x^+} e^{i(\xi+ \vp )}
\eea
where $\vp$ and $\xi$ are real parameters. Then the parameters
$a,b$ of the unitary representation are
\bea
a = \sqrt{g}\sqrt{i x^- - i x^+} e^{i( \xi+
\vp )} = \sqrt{g} \eta\,,\quad b=-\sqrt{g}{\sqrt{i x^- - i x^+}\ov
x^-} e^{i(\xi-\vp )}\,,
\eea
and the central charges are given by
\begin{eqnarray}\la{Hp}
H(p) = \sqrt{1+ 16 g^2\sin^2({1\ov 2}p)}\,,\quad C(p,\xi) =
ige^{2i\xi}(e^{ip}-1) \, .\end{eqnarray} The phase of the
parameter $\eta$ has been chosen so that $\vp =0$ would correspond
to the symmetric string theory choice we have discussed above.
% between the two $\su(2)$ subalgebras in $\su(2|2)$.
We will see in next sections that the S-matrix corresponding to
the symmetric choice satisfies the usual YB equation, and its
near-plane-wave expansion completely agrees with  the
leading-order string S-matrix computed in \cite{KMRZ}. With this
choice the ZF algebra for strings in $\AdS$ takes the usual form
too.

On the other hand, if one makes the spin chain choice $\vp = -\xi$
\cite{B} then the corresponding spin chain S-matrix satisfies the
twisted YB equation, and the ZF algebra has a more complicated
structure which involves a twist operator depending on the
world-sheet momentum $\hP$.

%%%%%%%%%%%%%%%%%%%%%%%%%%%%%%%%%%%%%%
\section{The $\su(2|2)$-invariant S-matrix}\la{sm}
Since the manifest symmetry algebra of the light-cone string
theory on $\AdS$ consists of two copies of the centrally-extended
$\su(2|2)$, the creation operators $A^\dagger_{M\dM}(p)$ carry two
indices $M$ and $\dot{M}$, where the dotted index is for the
second $\su(2|2)$.  The $n$-particle states are obtained by acting
by the creation operators on the vacuum
\begin{equation}
A^\dagger_{M_1\dM_1}(p_1)\cdots  A^\dagger_{M_n\dM_n}(p_n)|0 \rangle \equiv |A^\dagger_{M_1\dM_1}(p_1)\cdots  A^\dagger_{M_n\dM_n}(p_n) \rangle \,.
\end{equation}
For the purpose of this section we can think of  $A^\dagger_{M\dM}(p)$ as being a product of two creation operators $A^\dagger_{M\dM}(p)=A^\dagger_{M}(p)\times A^\dagger_{\dM}(p)$ and restrict our attention to the states created by $A^\dagger_{M}(p)$.

\subsection{Two-particle states and the S-matrix}
It is clear that a one-particle state $|A^\dagger_{M}(p)\rangle$ is identified with the basis vector $|e_M\rangle$ of the  fundamental representation $V(p,1)$ of
$\su(2|2)_{H,C}$. Let us stress that we have to set the parameter $\zeta$ to 1, because we use the canonical form of the central charge $\bC$ with $\xi=0$
\bea\la{Ccc}
\bC=ig\,(e^{i\hP}-1)\,,\quad \bC|A^\dagger_{M}(p)\rangle = ig\,(e^{ip}-1) |A^\dagger_{M}(p)\rangle\, .
\eea
We also set $\vp=0$ which for one-particle states describes both the string theory choice $\vp=0$, and  the spin chain choice $\vp = -\xi$.

Then the two-particle states created by $A^\dagger_{M}(p)$  should
be identified with the tensor product of fundamental
representations of $\su(2|2)_{H,C}$ \bea\la{tp}
|A^\dagger_{M_1}(p_1) A^\dagger_{M_2}(p_2) \rangle \sim
V(p_1,\zeta_1)\otimes V(p_2,\zeta_2)\,, \eea equipped with the
canonical action of the symmetry generators in the tensor product.
An important observation is that the parameters $\zeta_k$ cannot
be equal to 1 \cite{Bn}. The reason for that is very simple.
Computing the central charge $\bC$ on the two-particle state, we
get \bea\la{Ccn2} \bC |A^\dagger_{M_1}(p_1) A^\dagger_{M_2}(p_2)
\rangle=ig(e^{i(p_1+ p_2)}-1)|A^\dagger_{M_1}(p_1)
A^\dagger_{M_2}(p_2) \rangle\, , \eea because $\hP
A^\dagger_{M}(p) = A^\dagger_{M}(p) ( \hP + p)$. On the other hand
the value of the central charge on the tensor product of
fundamental representations is just equal to the sum of their
charges \bea \bC\, V(p_1,\zeta_1)\otimes V(p_2,\zeta_2) = ig\left(
\zeta_1(e^{ip_1}-1) + \zeta_2(e^{ip_2}-1) \right)
V(p_1,\zeta_1)\otimes  V(p_2,\zeta_2)\,.~~~ \eea Thus, we must
have the following identity \bea\la{iden2} e^{i(p_1+ p_2)}-1 =
\zeta_1(e^{ip_1}-1) + \zeta_2(e^{ip_2}-1)\,, \eea and it cannot be
satisfied if both $\zeta_1$  and $\zeta_2$ are equal to 1. In
fact, it is easy to show that there are only two solutions to this
equation for $\zeta_k$ lying on the unit circle \bea\la{z1z2} \{
\zeta_1 = e^{ip_2}\,,\  \zeta_2=1\}\,,\quad {\rm or} \quad  \{
\zeta_1 = 1\,,\  \zeta_2=e^{ip_1}\}\,. \eea Both choices can be
used  to identify a two-particle state with the tensor product. It
is readily seen that the first choice corresponds to the following
rearrangement of the commutation relation of the central charge
$\bC$ with $A^\dagger_M(p)$ \bea\la{ca1} \bC\, A^\dagger_M(p)=
C(p)A^\dagger_M(p)\,e^{i\hP} + A^\dagger_M(p)\,\bC\,, \eea which
is exactly of the form (\ref{JAs}). The second choice, on the
other hand, corresponds
 to another rearrangement of
the commutation relation
\bea\la{ca2}
\bC\, A^\dagger_M(p)= C(p)A^\dagger_M(p) +e^{ip}A^\dagger_M(p)\,\bC\,,
\eea
which is of the form (\ref{JAb}).

In general there is no restriction on the choice of the parameters
$\vp_i$ (or $\eta_i$) of the fundamental representations appearing in (\ref{tp}).
However, in what follows we will be interested in the string theory choice $\vp=0$, and in the spin chain choice $\vp = -\xi$.  Both choices describe equivalent physical theories, but the string theory choice leads to the S-matrix satisfying the usual YB equation, and as a result to the conventional ZF algebra.

Thus, making the first choice in (\ref{z1z2}), we see that the  invariance condition (\ref{incon2}) for the $\su(2|2)$ S-matrix takes the following form
for bosonic generators $L_a^b$ and $R_\a^\b$
\bea
S_{12}(p_1,p_2)\big(J \otimes \mathbb{I} + \mathbb{I} \otimes J\big)=\big( J \otimes \mathbb{I} + \mathbb{I} \otimes J\big)  S_{12}(p_1,p_2)\,, \la{incondb}\eea
and for fermionic generators $Q_\a^a$  and $\overline{Q}_a^\a$
\bea\nonumber
&&S_{12}(p_1,p_2)\big(J(p_1;e^{ip_2},\vp_1)\otimes \mathbb{I} + \Sigma\otimes J(p_2;1,\vp_2)\big)=\\
&&~~~~~~~~~~~~~~~~~~~~~~~\big( J(p_1;1,\tilde{\vp}_1)\otimes
\Sigma + \mathbb{I}\otimes J(p_2;e^{ip_1},\tilde{\vp}_2) \big)
S_{12}(p_1,p_2)\,,~~~\la{incondf} \eea where $J(p;\zeta,\vp)$
denote the structure constants matrices of the fundamental
representation parametrized by $p,\ \zeta = e^{2i\xi}$ and $\vp$,
see (\ref{repg}) and (\ref{abcd}), and  $\Sigma$ is the diagonal
matrix which takes care of the negative sign for fermions \bea
\Sigma = {\rm diag} (1,1,-1,-1)\,. \eea These are the conditions
that should be used to find the S-matrix. For the string symmetric
choice leading to the canonical S-matrix satisfying the YB equation we choose
 \bea\la{stb}
\mbox{\sc String\ theory\
basis}:\ \qquad\vp_1 =
\vp_2=\tilde{\vp}_1=\tilde{\vp}_2=0\,,~~~~~~~~~~~~~~~~~~~~~~~~~~~~~~~~~~~
\eea
and for the
spin chain choice we have
\bea\la{scb}\mbox{\sc Spin\ chain\
basis:\ }\qquad~~~~~~~~\vp_2 = \tilde{\vp}_1=0\,,\quad \vp_1=-{p_2\ov
2}\,,\quad \tilde{\vp}_2=-{p_1\ov 2}\,.~~~~~~~~~~~~~~~~~~~~~~
\eea

The form of the structure constants matrices $J(p;\zeta,\vp)$
allows us to
 determine the commutation relations (\ref{JAs})  of the symmetry operators with the creation and annihilation operators. It is convenient to use the matrix notations, i.e. to combine $A^\dagger_M$ and $A_M$ into a row and column, respectively, and the symmetry algebra structure constants into matrices $L_a^b$, $R_\a^\b$, $Q_\a^a$ and $\overline{Q}_a^\a$, see (\ref{repg}).
 Then, for the string theory choice of the basis $\vp_i=0$,  the relations
(\ref{JAs}) for the centrally-extended algebra $\su(2|2)$  can be
written in the following simple form \bea\la{JAs2} \mbox{\sc
String\ theory\ basis:}~~~~ &&\bL_a{}^b A^\dagger(p)
=A^\dagger(p)\,L_a^b+ A^\dagger(p)\,\bL_a{}^b
\,,~~~~\\\nonumber &&\bR_\a{}^\b A^\dagger(p)
=A^\dagger(p)\,R_\a^\b+ A^\dagger(p)\,\bR_\a{}^\b \,,\\\nonumber
&& \bQ_\a{}^a A^\dagger(p)
=A^\dagger(p)\,Q_\a^a(p)\, e^{i\hP/2}+A^\dagger(p)\,\Sigma\,
\bQ_\a{}^a \,,\\\nonumber
&& \bQ_a^\dagger{}^\a A^\dagger(p)
=A^\dagger(p)\,\overline{Q}_a^\a(p)\,
e^{-i\hP/2}+A^\dagger(p)\,\Sigma\, \bQ_a^\dagger{}^\a
\,,\nonumber ~~~~~~~~~~~\eea
where $L_a^b$, $R_\a^\b$, $Q_\a^a$ and $\overline{Q}_a^\a$ are the  symmetry algebra structure constants matrices of the one-particle representation with $\xi=\vp=0$.
Thus, for the string theory basis the braiding factors in (\ref{JAs}) are just scalar operators $e^{\pm i\hP/2}$.

In the case of the spin chain choice of the basis (\ref{scb}), the relations (\ref{JAs}) take the following more complicated form
 \bea\la{JAs3}\mbox{\sc Spin\ chain\ basis:}~~~~ &&\bL_a{}^b A^\dagger(p)
=A^\dagger(p)\,L_a^b+ A^\dagger(p)\,\bL_a{}^b \,,\\\nonumber
&& \bR_\a{}^\b A^\dagger(p) =A^\dagger(p)\,R_\a^\b+
A^\dagger(p)\,\bR_\a{}^\b \,,\\\nonumber && \bQ_\a{}^a A^\dagger(p)
= A^\dagger(p)\,Q_\a^a(p)\, \Theta(\hP)+A^\dagger(p)\,\Sigma\,
\bQ_\a{}^a \,,\\\nonumber && \bQ_a^\dagger{}^\a A^\dagger(p)
= A^\dagger(p)\,\overline{Q}_a^\a(p)\,
\overline{\Theta}(\hP)+A^\dagger(p)\,\Sigma\, \bQ_a^\dagger{}^\a
\,,\nonumber~~~~~~~~~~~~~~~~~~
\eea
 where  the braiding factors
$\Theta(\hP)$ and $\overline{\Theta}(\hP)$ are the diagonal
matrices \bea\la{thetas} \Theta(\hP)= {\rm diag}
(1,1,e^{i\hP},e^{i\hP})\,,\qquad \overline{\Theta}(\hP) =
e^{-i\hP}\Theta(\hP)\,. \eea
It is not difficult to check that the commutation relations (\ref{JAs3}) follow from (\ref{JAs2}) after the following change of the creation/annihilation operators
\bea\la{utaa}
A^\dagger(p) \to A^\dagger(p)\, U(\hP)\,,\quad A(p) \to U^\dagger(\hP) A(p)\,,
\eea
where
\bea\la{Umo}
U(\hP)= {\mbox {diag}} ( e^{{i\ov 2}\hP}, e^{{i\ov 2}\hP}, 1,1)\,.
\eea
Let us stress that the transformation (\ref{utaa}) is not unitary because as we will see soon it changes the commutation relation of the creation and annihilation operators, and, therefore, the form of the ZF algebra.

\medskip

It is worthwhile to notice that  the form of the two-particle structure constants matrices
appearing in the invariance condition (\ref{incondf}) allows us to reformulate the problem by using the Hopf algebra language similar to the considerations in \cite{Gomez:2006va}, see appendix (\ref{hopf}) for detail.

\medskip

It is also interesting  to mention that the (anti)-commutation relations (\ref{JAs2}) (and (\ref{JAs3})) of the symmetry algebra generators with the creation and annihilation operators are written in the usual (anti)-commutator form. The only difference with the standard relations is in the appearance of the operator $e^{\pm i\hP/2}$ on the r.h.s. of the relations  (\ref{JAs2}).  It is easy to see that one can get rid of the dependence on  $e^{\pm i\hP/2}$ in these relations by redefining the supersymmetry generators
$\bQ_\a{}^a$ and $\bQ_a^\dagger{}^\a$ as follows
\bea\la{redq}
\bQ_a{}^\a\to \bQ_a{}^\a\,  e^{i\hP/2} \,,\quad
\bQ_a^\dagger{}^\a\to \bQ_a^\dagger{}^\a\, e^{-i\hP/2}\,.
\eea
Then the relations  (\ref{JAs2}) for the redefined supersymmetry charges take the form of the braided (anti)-commutators
\bea\la{JAs2b}
 \bQ_\a{}^a A^\dagger(p)-e^{-ip/2}A^\dagger(p)\,\Sigma\, \bQ_\a{}^a
&=&A^\dagger(p)\,Q_\a^a(p)\, e^{-ip/2} \,,\\\nonumber
\bQ_a^\dagger{}^\a A^\dagger(p)- e^{ip/2}A^\dagger(p)\,\Sigma\, \bQ_a^\dagger{}^\a
&=&A^\dagger(p)\,\overline{Q}_a^\a(p)\,
e^{ip/2}
\,.\eea
It is the form one usually discusses considering models with nonlocal charges
\cite{Bernard:1990ys}. The redefinition (\ref{redq}) changes the $\hP$-dependence of the central charge $\bC$:
\bea
\bC \to ig\,(e^{i\hP}-1) e^{-i\hP} =  ig\,(1 - e^{-i\hP})\,,
\eea
but it does not change the form of the S-matrix if one keeps the track of the additional phases. We will not be using this form of the commutation relations in this paper.

\medskip

Finally, let us note that, if we think of vectors from $V(p;\zeta)$ as
columns then, as can be seen from the formula (\ref{incondf}), the
$S$-matrix can be considered as the map \bea S_{12}(p_1,p_2):
~~~~~V(p_1,e^{ip_2})\otimes V(p_2,1)\to V(p_1,1)\otimes
V(p_2,e^{ip_1}) \,, \eea and if we think of vectors from $V(p;\zeta)$
as rows then the $S$-matrix can be considered as the opposite map \bea
S_{12}(p_1,p_2): ~~~~~V(p_1,1)\otimes V(p_2,e^{ip_1}) \to
V(p_1,e^{ip_2})\otimes V(p_2,1) \,.  \eea From this point of view the
action of the S-matrix corresponds to exchanging the two possible
choices of the parameters $\zeta_k$ of the two representations.  Let
us stress however, that no matter what interpretation we use,
$S_{12}(p_1,p_2)$ is a $16\times 16$ matrix acting in the
16-dimensional vector space of the two-particle states
$|A^\dagger_{M_2}(p_2) A^\dagger_{M_1}(p_1) \rangle$, and as we
discussed in section 2, if $p_1=p_2$ the S-matrix reduces to the minus
permutation.\footnote{Due to the definition of the S-matrix we are
  using, if $p_1=p_2$ it is the minus permutation but not the graded
  permutation.}  Then, if the string coupling constant $g$ goes to
infinity the string sigma-model becomes free, and the ZF creation
operators become the usual creation operators and just commute or
anti-commute depending on the statistics, and, therefore, in this
limit the S-matrix must be equal to the graded unity.

The S-matrix satisfying the invariance conditions (\ref{incondb}) and (\ref{incondf}) can be easily found up to an overall scalar factor, and its explicit form is given in Appendix \ref{appsmat},
eq.(\ref{Smatrix}).

\subsection{Multi-particle states}

Multi-particle states created by $A^\dagger_{M}(p)$  are
correspondingly identified with the tensor product of fundamental
representations of $\su(2|2)_{H,C}$ \bea\la{tpm}
|A^\dagger_{M_1}(p_1)\cdots  A^\dagger_{M_n}(p_n) \rangle \sim
V(p_1,\zeta_1)\otimes \cdots \otimes V(p_n,\zeta_n)\,, \eea
equipped with the canonical action of the symmetry generators in
the tensor product, and the parameters  $\zeta_k$ have to satisfy
the following identity \bea\la{iden} e^{i(p_1+\cdots + p_n)}-1 =
\sum_{k=1}^n\zeta_k(e^{ip_k}-1)\,. \eea In general, there are many
different solutions to this equation. In our case, however, the
choice of $\zeta_k$ is fixed by the commutation relations
(\ref{JAs}) and (\ref{ca1}). One can easily see that the only
solution compatible with (\ref{JAs}) is \bea
\zeta_1=e^{i(p_2+\cdots +p_n)}\,,\quad \zeta_2=e^{i(p_3+\cdots
+p_n)}\,,\quad \ldots\,, \quad \zeta_{n-1}=e^{ip_n}\,,\quad
\zeta_n=1\,.~~~ \eea On the other hand the only solution
compatible with (\ref{JAb}) and (\ref{ca2})  is \bea
\zeta_1=1\,,\quad \zeta_2=e^{ip_1}\,,\quad \ldots\,, \quad
\zeta_{n-1}=e^{i(p_1+\cdots +p_{n-2})}\,,\quad
\zeta_n=e^{i(p_1+\cdots +p_{n-1})}\,.~~~ \eea The multi-particle
S-matrix just maps the vector space with the first choice of
$\zeta_k$ to the (isomorphic) space with the second choice of
$\zeta_k$. For generic values of the parameters $\vp_i$ the
factorizability condition for the S-matrix appears to be
equivalent to a version of the YB equation which follows from a
twisted Zamolodchikov algebra. The string theory choice $\vp_i=0$
is singled out because the corresponding string S-matrix satisfies
the usual YB equation, and, therefore, can be used to construct
the usual ZF algebra.

%%%%%%%%%%%%%%%%%%%%%%%%%%%%%%%%%%%%%%%%%%
\section{The Zamolodchikov-Faddeev  algebra and the twisting}
\subsection{The Zamolodchikov-Faddeev algebra}
The Zamolodchikov-Faddeev algebra for strings in $\AdS$ involves
creation and annihilation operators which carry two indices $M$
and $\dot{M}$ corresponding to the two centrally-extended
$\su(2|2)$ subalgebras in the symmetry algebra of the light-cone
string theory. In what follows we use the indices $i,j,k,...$ to
denote the pairs of indices $(M,\dM)$, $(N,\dN),...$, and use the
notations from Appendix \ref{notations}.

The $\AdS$ string S-matrix that describes the
scattering of two-particle states is a tensor product of two
$\su(2|2)$ S-matrices \bea \S_{ij}^{kl}(p_1,p_2)\equiv
\S_{M\dM,N\dN}^{K\dK,L\dL}(p_1,p_2) = \S_{MN}^{KL}(p_1,p_2)
\S_{\dM\dN}^{\dK\dL} (p_1,p_2)\,. \eea
Here the $\su(2|2)$ invariant S-matrix $\S_{MN}^{KL}(p_1,p_2)$  includes the necessary scalar factor, see appendix \ref{appsmat}
\bea\la{smatstr}
\S_{MN}^{KL}(p_1,p_2) = S_0(p_1,p_2)S_{MN}^{KL}(p_1,p_2)\,,
\eea
and we use the canonical S-matrix corresponding to the string symmetric choice $\vp_i=0$, because only under this choice of the parameters (up to trivial transformations) the S-matrix satisfies the normal YB equation.

Let us stress that even though the S-matrix is the tensor product
the ZF algebra  can be formulated only in terms of the two-index operators
 $A^\dagger_{M\dM}(p)$ and $A^{M\dM}(p)$  because
it contains their commutation relations.
One cannot think of the creation and annihilation operators as the products
$A^\dagger_{M\dM}(p) =A^\dagger_{M}(p)A^\dagger_{\dM}(p)$ and $A^{M\dM}(p)=A_{M}(p)A_{\dM}(p)$ with the one-index operators forming independent $\su(2|2)$ ZF algebras.\footnote{However,
all consistency conditions for the full S-matrix just follow from the same conditions for the $\su(2|2)$ invariant S-matrix $\S_{MN}^{KL}(p_1,p_2)$.}

The ZF algebra has the form
\bea\la{TZF}
 A_1A_2&=&
\S_{12}A_2A_1\,,\quad A^\dagger_1A^\dagger_2=
A^\dagger_2A^\dagger_1\S_{12}\,,\quad A_1A^\dagger_2=
A^\dagger_2\S_{21}A_1 + \delta_{12}\,. \eea Here we use the
standard matrix notations, see Appendix \ref{notations} for
details, in particular \bea\nonumber A^\dagger_1A^\dagger_2&=&
\sum_{i,j} A^\dagger_i(p_1)A^\dagger_j(p_2)\,  E^i \otimes
E^j\,,\quad A^\dagger_2A^\dagger_1= \sum_{i,j}
A^\dagger_j(p_2)A^\dagger_i(p_1)\,  E^i\otimes E^j \\\nonumber
\S_{12}&=& \sum_{ijkl} \S_{ij}^{kl}(p_1,p_2)\,  E^i_k \otimes
E^j_l\,, \quad\quad  \delta_{12}= \delta(p_1-p_2) \sum_i
E_i\otimes E^i\,. \eea

Let us now recall the consistency conditions for the
ZF algebra. The first one is the unitarity condition
for the S-matrix
 \bea\la{uc}
 \S_{12}(p_1,p_2)\S_{21}(p_2,p_1)= \mathbb{I}\,,
\eea which follows by applying the commutation relation
$A^\dagger_1A^\dagger_2= A^\dagger_2A^\dagger_1\S_{12}$ twice.
Then, the ZF algebra should have a unique basis of the
lexicographycally ordered monomials (i.e. it has the
Poincar\'e-Birkhoff-Witt property) and no new relations except
(\ref{TZF}) arise. This requirement leads to  the YB equation. By
using the two different ways of reordering
$A^\dagger_1A^\dagger_2A^\dagger_3$ to
$A^\dagger_3A^\dagger_2A^\dagger_1$, we get \bea\nonumber
&&A^\dagger_1A^\dagger_2A^\dagger_3=A^\dagger_3A^\dagger_2A^\dagger_1
\S_{12}\S_{13}\S_{23}\,,\\\nonumber
&&A^\dagger_1A^\dagger_2A^\dagger_3=A^\dagger_3A^\dagger_2A^\dagger_1
\S_{23}\S_{13}\S_{12}\,, \eea and, therefore, derive the YB
equation \bea\la{YBnorm}
\S_{23}(p_2,p_3)\S_{13}(p_1,p_3)\S_{12}(p_1,p_2)
=\S_{12}(p_1,p_2)\S_{13}(p_1,p_3)\S_{23}(p_2,p_3)\,.  \eea Let us
mention that both the l.h.s. and r.h.s. of this equation represent
the 3-particle scattering S-matrix, and the equation itself is the
factorizability condition for the S-matrix. Realizing the
$\su(2|2)$ invariant S-matrix (\ref{Smatrix}) as a $16\times 16$
matrix it is a straightforward exercise to check  that it
satisfies both the unitarity condition (\ref{uc}) and the YB
equation (\ref{YBnorm}).

\medskip

The ZF algebra also satisfies the physical unitarity condition that follows from the fact  that
the annihilation operators are hermitian conjugate of the creation operators.  To
derive  the condition of the physical unitarity we find instructive to use explicit index notations.
The commutation relations for the creation operators are of the form
$$
A_i^\dagger(p_1)A_j^\dagger(p_2)=A^\dagger_{l}(p_2)A^\dagger_k(p_1)\S_{ij}^{kl}(p_1,p_2)\,.
$$
Taking into account  $(A_i^\dagger(p))^{\dagger}=A^i(p)$, we conjugate the relation above and change
$p_1\leftrightarrow p_2\,,\ i\leftrightarrow j\,,\ k\leftrightarrow l$ . Then we get the following commutation relations for the annihilation operators
$$
A^i(p_1)A^j(p_2)=\S_{ji}^{* lk}(p_2,p_1)A^l(p_2)A^k(p_1)
$$
According to our assumptions about the structure of the ZF algebra
it must be equal to
$$
A^i(p_1)A^j(p_2)=\S_{kl}^{ij}(p_1,p_2)A^l(p_2)A^k(p_1)\,.
$$
Thus, the S-matrix must satisfy the following identity
$$
\S_{ji}^{* lk}(p_2,p_1) = \S_{kl}^{ij}(p_1,p_2)\,,
$$

which can be rewritten in the matrix form
\bea\la{ucp}
\S_{21}^{\dagger}(p_2,p_1)=\S_{12}(p_1,p_2)\, .
\eea
This is the condition of the physical initarity.
Taking into account the usual unitarity condition for $S$-matrix (\ref{uc}),
 we can write the physical unitarity condition (\ref{ucp})
of the S-matrix in the following form:
$$\S^{\dagger}(p_1,p_2)\S(p_1,p_2)={\mathbb I}.$$ It is
 easy to check
numerically that this condition holds for the $\su(2|2)$ invariant
S-matrix (\ref{Smatrix}) with $\eta$'s satisfying the unitary
representation conditions (\ref{urep}).

\medskip

An important property of the ZF algebra is that it possesses an
abelian subalgebra generated by the commuting operators of the
form \bea \bI_\om = \int\,{\rm d}p\, \om(p)
A_i^\dagger(p)A^i(p)\,, \eea where $\om(p)$ is an arbitrary
function. In particular, the world-sheet momentum $\hP$ and the
Hamiltonian $\bH$ belong to this subalgebra \bea\la{bbH} \hP =
\int\,{\rm d}p\, p\, A_i^\dagger(p)A^i(p)\,,\quad  \bH =
\int\,{\rm d}p\, H(p) A_i^\dagger(p)A^i(p)\,, \eea where $H(p)$ is
the value of the Hamiltonian on a one-particle state with momentum
$p$: $H(p) = \sqrt{1+ 16 g^2\sin^2({1\ov 2}p)}$. It is not
difficult to check that 
\bea 
\bI_\om\, A_i^\dagger(p)=A_i^\dagger(p)\left( \om(p) + \bI_\om \right) \ 
\eea
leading to the additivity property of the commuting integrals \bea
\bI_\om\,A^\dagger_{i_1}(p_1)\cdots A^\dagger_{i_n}(p_n)|0 \rangle
=\left(\sum_{k=1}^n\om(p_{i_k})\right) A^\dagger_{i_1}(p_1)\cdots
A^\dagger_{i_n}(p_n)|0 \rangle\,. \eea The existence of the family
of the commuting integrals together with the additivity property
of the integrals guarantees the integrability of the model.

%%%%%%%%%%%%%%%%%%%%%%%%%%%%%
\subsection{Twisting the ZF algebra}
%%%%%%%%%%%%%%%%%%%%%%%%%%%%%%
As was mentioned in the previous section,  the spin chain S-matrix corresponding to the choice  $\vp=-\xi$ in
(\ref{urep}) does not satisfy the usual YB equation, and therefore cannot be used to construct the standard ZF algebra.
In this subsection we show that the spin chain S-matrix appears in
a twisted ZF algebra of the form (\ref{ZFg}) in such a way which
leads to a modified YB equation satisfied by the S-matrix.
The twisted Zamolodchikov algebra with
the spin chain S-matrix
is equivalent to the ZF algebra with the string theory S-matrix because they are
related by a transformation of the creation/annihilation operators.

As we discussed in section \ref{Repr}, the parameter $\eta$ just reflects our freedom in the choice of the basis of a fundamental representation. On the ZF algebra level this freedom is implemented by the following transformation of the creation/annihilation operators
\bea\la{uta}
A^\dagger(p) \to A^\dagger(p)\, U(\hP, p)\,,\quad A(p) \to U^\dagger(\hP,p) A(p)\,,
\eea
where $A^\dagger$ is a row, $A$ is a column, and $U$ is an arbitrary unitary matrix which can depend on the momentum $p$ of the
creation/annihilation operators, and on the world-sheet momentum operator $\hP$. This transformation is not an automorphism of the ZF algebra. It is not difficult to see that under this transformation the  ZF algebra can be again written in the same form (\ref{TZF}) but with the following transformed operator-valued S-matrix
\bea\la{soper}
\S_{12}^{U}(p_1,p_2; \hP)= U_2(\hP +p_1,p_2) U_1(\hP, p_1)\S_{12}(p_1,p_2) U_2^\dagger(\hP,p_2)U_1^\dagger(\hP+p_2,p_1)\,.~~~~~~
\eea
Obviously, the  twisted ZF algebra is isomorphic to the original ZF one, and they describe one and the same physical system.   The corresponding two-particle S-matrix $\S_{12}$ transforms, however, in a nontrivial way
\bea\la{tS2}
\S_{12}^U(p_1,p_2)= U_2(p_1,p_2) U_1(0, p_1)\S_{12}(p_1,p_2) U_2^\dagger(0,p_2)U_1^\dagger(p_2,p_1)\,,~~~~~~
\eea
where we have taken into account that $\hP|0\rangle =0$.  We see, therefore, that such a transformation allows one to introduce a number of unphysical parameters and spurious dependence of momenta $p_i$. It is clear that the transformed S-matrix $S^U_{12}$ satisfies twisted YB equation with the twist determined by the matrix $U$.

To simplify the notations in the rest of this subsection we only consider the Zamolodchikov subalgebra with the $\su(2|2)$ invariant S-matrix   generated by the creation operators $A^\dagger_M(p)$. The generalization to the full ZF algebra with the string S-matrix is straightforward.

The transformation (\ref{uta}) can, in particular,  be used to
change the parameters $\eta_i$ which appear in the two-particle
S-matrix. For example, it is clear from the discussion in section
\ref{sm}, see (\ref{utaa}), that to obtain the spin chain S-matrix
corresponding to the  choice of $\eta$ with $\vp=-\xi$, see
(\ref{urep}) one should consider a diagonal matrix $U$ of the
following form \bea\la{Um} U(\hP, p)\equiv U(\hP)= {\mbox {diag}}
( e^{{i\ov 2}\hP}, e^{{i\ov 2}\hP}, 1,1)\,. \eea Note, that $U$
has no dependence on $p$ because we do not want to produce a
trivial momentum dependent phase for $\eta$.  To see that this
transformation produces the desired effect we consider the
corresponding change of a two-particle state \bea\la{taa}
A_1^\dagger(p_1)A_2^\dagger(p_2)|0\rangle \to
A_1^\dagger(p_1)U_1(p_2)A_2^\dagger(p_2)|0\rangle\,. \eea As we
discussed in section \ref{sm}, the two-particle state is
identified with the tensor product of the fundamental
representations $V(p_1,e^{ip_2})\otimes V(p_2,1)$. We see,
therefore, that the transformation (\ref{taa}) implies a change of
the basis of the vector space $V(p_1,e^{ip_2})$. One can easily
check that this change means a multiplication of the constant $a$
in (\ref{repg}) by the phase $e^{-{i\ov 2}p_2}$ and, since
$\xi_1={p_2\ov 2}$, this leads to the spin chain choice of
$\eta_1$ with the phase of $\eta_1$ equal to $0$. There is no need
to perform any change of the basis of the space $V(p_2,1)$ because
for this representation $\vp_2 = \xi_2=0$. Finally, eq.(\ref{tS2})
which relates the spin chain S-matrix  to the canonical string
S-matrix (\ref{Smatrix}) with $\vp=-\xi$ takes the following form
\bea\la{tS3} S_{12}^{\rm chain}(p_1,p_2)= U_2(p_1) S_{12}^{\rm
string}(p_1,p_2) U_1^\dagger(p_2)\,.~~~~~~ \eea Taking into
account eq.(\ref{tS3}), we see that the operator-valued S-matrix
(\ref{soper}) can be cast in the following form \bea\la{soper2}
S_{12}^{\rm chain}(p_1,p_2; \hP)= F_{12}(\hP)S_{12}^{\rm
chain}(p_1,p_2)F_{12}^{-1}(\hP) \,,~~~~~~ \eea where $F_{12} =
U(\hP)\otimes U(\hP)$ is  the twist matrix. The twisted
Zamolodchikov algebra then takes the form \bea\la{TZ}
A^\dagger_1A^\dagger_2= A^\dagger_2A^\dagger_1S_{12}^{\rm
chain}(p_1,p_2; \hP)\,,\quad \hP A^\dagger_1 = A^\dagger_1( \hP +
p_1)  \,, \eea leading to the following twisted YB
equation\footnote{The S-matrix constructed in the gauge theory
framework in \cite{B} is an operator S-matrix because besides
being a matrix it also operates by inserting $Z^{\pm}$ symbols in
an infinite chain of $Z$ fields. As such, it satisfies the usual
YB equation since the twisting is compensated by the effect
produced by insertions. In our approach, however, we never deal
with operator S-matrices.} for $S_{12}^{\rm chain}(p_1,p_2)$
\bea\la{TYB}
 F_{23}(p_1) S_{23}F_{23}^{-1}(p_1)S_{13}F_{12}(p_3)S_{12}F^{-1}_{12}(p_3)
=S_{12}F_{13}( p_2)S_{13}F_{13}^{-1}(p_2)S_{23}\,,  \eea where
$S_{ij}\equiv S_{ij}^{\rm chain}(p_i,p_j)$. Let us finally mention
that, just as in the case of the usual ZF algebra, both the lhs
and rhs of this equation represent the 3-particle scattering
S-matrix, and the equation itself is the factorizability condition
for the S-matrix. It is clear, however, that it is the S-matrix
satisfying the standard YB equation that should be considered as
the \emph{canonical} $\AdS$ string S-matrix.

%%%%%%%%%%%%%%%%%%%%%%%%%%
\section{Crossing symmetry}
It is well known that the S-matrix in relativistic quantum
integrable systems satisfies an additional property called
crossing symmetry.
%It arises upon substituting in the scattering
%process one of the particles for its antiparticle.
%In particular,
%crossing symmetry
This relation allows one to express the S-matrix for the
particle-antiparticle scattering via the S-matrix for scattering of
two particles.  From the point of view of the ZF algebra the crossing
symmetry can be regarded as a certain algebra automorphism.  \medskip

Though our string model does not possess the relativistic invariance on
the world-sheet, it is still reasonable to require that
the corresponding  ZF algebra has an additional invariance
related to the particle-antiparticle symmetry of the theory. More
precisely, we define the particle-to-antiparticle transformation as
 \bea\la{pat} A^\dagger(p)\to B^\dagger(p)= A^{t}(-p)
\mathscr{C}(-p)\,,\quad A(p)\to B(p)=\mathscr{C}^{\dagger}(-p)
A^\dagger{}^{t}(-p) \,,~~~~ \eea where $\mathscr{C}(p)$ is a
``charge-conjugation'' matrix which a priori may  depend on $p$,
and superscript $t$ means the usual matrix transposition. We
require this map to be an automorphism of the  ZF algebra
for $p_1\neq p_2$. This means that if we first replace in the
algebra relations $A$ by $B$, and further use the formulas
(\ref{pat}) to express $B$ via $A$ we should recover for $A$ the
same ZF algebra. Under the usual assumption $p_1>p_2$ the
delta-function does not contribute which makes it possible to map
by using (\ref{pat}) the exchange relations of $A(p_1)$ and
$A(p_2)$ to that of $A(p_1)$ and $A^\dagger(p_2)$. Obviously, the
requirement of the crossing symmetry should impose certain
restrictions both on $\mathscr{C}(p)$ and on the S-matrix.

\medskip
Note that flipping the sign of $p$ under the
particle-to-antiparticle transformation (\ref{pat}) is also an
immediate consequence of the fact that this map is an automorphism
of the  ZF algebra. Indeed, this change of the sign is
dictated by the compatibility of the particle-to-antiparticle
transformation with the algebra relations
$$
\hP A^\dagger = A^\dagger( \hP + p)  \,,\quad \hP  A = A( \hP -
p)\,,
$$
and the requirement that the symmetry algebra generators remain
untouched by the transformation (\ref{pat}).

\subsection{The
particle-to-antiparticle transformation} To understand the implications of
crossing symmetry we first note that the exchange relations
(\ref{JAs2}) should remain invariant under the
particle-to-antiparticle transformation (\ref{pat}), and this
imposes sever restrictions on the form of $\mathscr{C}$. From the
first two equations in (\ref{JAs2}) we find the following
relations \bea
\la{rel1}
&&\mathscr{C}(p)\,L_a^b=- L_b^a\, \mathscr{C}(p)\,,\\
\nonumber &&\mathscr{C}(p)\,R_\a^\b=-R_\b^\a\,
\mathscr{C}(p)\, , \eea where we take into account that
$(L_a^b)^\dagger = L_b^a$, and $(R_\a^\b)^\dagger =R_\b^\a$. These
relations fix the form of $\mathscr{C}$ up to two coefficients
$$ \mathscr{C}(p)=\left(\begin{array}{cccc} 0 & -c_1(p) & 0 & 0\\
c_1(p) & 0 & 0 & 0\\
0 & 0 & 0 & -c_2(p) \\
0 & 0 & c_2(p) & 0\\
 \end{array}\right)\, .
$$
The last two equations in (\ref{JAs2}) lead to the following
relations \bea
\la{rel2}
&&e^{i\frac{p}{2}}\mathscr{C}(p)\,\overline{Q}_a^\a(-p)=-
\left(\overline{Q}_a^\a(p) \right)^t\,
\Sigma\mathscr{C}(p)\,,\\
\nonumber &&\mathscr{C}(p)\,\,Q_\a^a(-p)=-
e^{i\frac{p}{2}}\left(Q_\a^a(p) \right)^t\, \Sigma\mathscr{C}(p)\
\, , \eea where we take into account that $(Q_\a^a(p))^\dagger
=\overline{Q}_a^\a(p)$. These relations not only  fix the form of
$\mathscr{C}$ but also determine how the coefficients $a,b,c,d$ of
the representation (\ref{repg}) transform under the change $p\to
-p$. We find
 \bea\la{cpp} &&\frac{c_2(p)}{c_1(p)}=
\frac{b(-p)}{a(p)}e^{\frac{i}{2}p}= \frac{2\sin{p\ov
2}}{\eta(p)\eta(-p)}\zeta(p)\,, \eea
 and
 \bea \la{abcdt}
&&{a(-p)\ov b(p)} = -{a(p)\ov b(-p)}e^{-i p} = {c(-p)\ov
d(p)}e^{-i p}=-{c(p)\ov d(-p)}\,. \eea It is not difficult to see
that the relations (\ref{abcdt}) imply that the transformed
coefficients satisfy the condition
$$
a(-p)d(-p)-b(-p)c(-p)=1\,,
$$
and, therefore, they define a fundamental representation of the
centrally-extended $\su(2|2)$. This is the antiparticle
representation, and one can read off the values of its central
charges
$$
H(-p) = - H(p)\,,\quad C(-p)= - C(p)e^{-ip}\,,\quad
\overline{C}(-p) =- \overline{C}(p)e^{ip}\,.
$$
In fact, it is straightforward to show that the relations
(\ref{abcdt}) give the following transformation law for the
coefficients $x^\pm$ and $\zeta$ under the
particle-to-antiparticle transformation (\ref{pat}) \bea x^\pm(-p)
= {1\ov x^\pm(p)}\,,\qquad \zeta(-p)=\zeta(p)\,. \eea Taking into
account that for the unitary representation with the symmetric
string theory choice $\eta(p) =
 \sqrt{i x^- - i x^+}\sqrt{\zeta}$, we
simply get \bea \frac{c_2(p)}{c_1(p)}={\sin {p\ov 2}\ov
\sqrt{-\sin^2 {p\ov 2}}}=- i\,{\rm sign}( p)\, ,
 \eea
 i.e the ratio of
the coefficients of the charge conjugation matrix is (almost)
momentum independent. Note that the map (\ref{pat}) reflects the
fact that $\su(2|2)$ has the structure of the
$\mathbb{Z}_4$-graded Lie algebra. Without loss of generality we can require the coefficient
$c_1$ to be independent on $p$ and satisfy $\bar{c}_1c_1=1$. Then
the matrix $\mathscr{C}$ takes the form
\bea\la{cc}
\mathscr{C}(p)=\left(\begin{array}{cc} \sigma_2 & 0 \\ 0 &- i\,{\rm sign}( p)\,
\sigma_2
\end{array}\right)\, ,
\eea
where $\sigma_2$ is the Pauli matrix. Obviously, successive
application of the map (\ref{pat}) four times gives the identity.
This is nothing else as a reflection of the fact that  $\su(2|2)$
admits a structure of the $\mathbb{Z}_4$-graded Lie superalgebra,
see e.g. \cite{Berkovits:1999zq}.

\subsection{Crossing symmetry condition}

As was discussed in section 4.3, the S-matrix appears to be
uniquely determined by symmetries up to a scalar factor which
we call $S_0(p_1,p_2)$. A possible functional form of
$S_0(p_1,p_2)$ is restricted by unitarity together with the
requirement of crossing symmetry \cite{Janik}. We therefore
write the S-matrix obeying the condition of crossing symmetry
in the form
$$
{\cal S}_{12}(p_1,p_2)=S_0(p_1,p_2)~ S_{12}(p_1,p_2)  \, ,
$$
where $S_{12}(p_1,p_2)$ is the S-matrix in the string basis; see
appendix \ref{appsmat}.

If we assume that $p_1>p_2$ then the map (\ref{pat}) is an
automorphism of the ZF algebra provided the matrix ${\cal
S}_{12}(p_1,p_2)$ obeys the following equations
\bea\la{cros}
&&
\mathscr{C}_1^{-1}(-p_1){\cal S}_{12}^{t_1}(p_1,p_2)
\mathscr{C}_1(-p_1)
 {\cal S}_{12}(-p_1,p_2)={\mathbb I}\, ,\\\nonumber
 &&
\mathscr{C}_2^{-1}(-p_2){\cal S}_{21}^{t_2}(p_2,p_1)
\mathscr{C}_2(-p_2)
 {\cal S}_{21}(-p_2,p_1)={\mathbb I}\, .
 \eea
Here $t_1$ and $t_2$ mean the transposition in the first and
second space, respectively, $\mathscr{C}_1 = \mathscr{C}\otimes
I$, $\mathscr{C}_2 = I\otimes \mathscr{C}$ and $\mathscr{C}$ is
the charge conjugation matrix (\ref{cc}). These two equations are
in fact equivalent because the first one turns into the second one
after applying the permutation and exchanging $p_1$ and $p_2$.

\medskip

Substituting here the string S-matrix from appendix \ref{appsmat}
we indeed find that relations (\ref{cros}) produce the following
equation for the scalar factor \bea S_0(-p_1,p_2)S_0(p_1,p_2)=
{1\ov f(p_1,p_2)}\, , \eea where the function  $f$ was introduced
in \cite{Janik} \bea
f(p_1,p_2)=\frac{\Big(\frac{1}{x_1^+}-x_2^-\Big)(x_1^+ -
x_2^+)}{\Big(\frac{1}{x_1^-}- x_2^- \Big)(x_1^--x_2^+)}\, .
\la{Janik} \eea Equation (\ref{Janik}) is precisely the same as
the one found by Janik \cite{Janik}. However, in opposite to
\cite{Janik}, our derivation does not refer to the $\su(1|2)$
invariant S-matrix, rather we use the $\su(2|2)$ S-matrix in the
string basis. Moreover, since $\mathscr{C}$ is the usual charge
conjugation matrix, the crossing equation (\ref{cros}) has the
usual form known in relativistic models. Also, to derive the
crossing equation we did not make any reference to a Hopf algebra
structure of the symmetry algebra.

Finally, we remark that the crossing equation for the scalar factor
 is incompatible with the assumption that  the S-matrix is an
analytical function of $p_1,p_2$. This is reflected by the
fact that the function $f(p_1,p_2)$ obeys the following properties
$$
f(p_2,p_1)f(-p_1,p_2)=1\, , ~~~~~f(p_2,p_1)f(p_1,-p_2)=1\, ,
$$
which are incompatible with the assumption of both unitarity and
analyticity.

In conclusion, we see that in the string basis the S-matrix  not
only satisfies the usual Yang-Baxter equation but it also admits a
scalar factor which allows for the standard crossing symmetry
condition.

\section{Comparison with the near plane-wave S-matrix}
In this section we compare the string theory S-matrix we used to
define the ZF algebra to the S-matrix  arising in the near
plane-wave limit of string theory on $\AdS$. The later was
recently computed in \cite{KMRZ} by reading off the quartic
interaction vertices of the string Lagrangian \cite{LCpaper}
obtained in the generalized uniform light-cone gauge\footnote{See
\cite{Callan}-\cite{We} on an extensive work concerning the
construction of the string Hamiltonian and finding near plane-wave
corrections to energies of the plane-wave states.}
\cite{We,magnon}. As the result, the corresponding S-matrix,
${\mathbb S}^{\rm KMRZ}$, appears to have an explicit dependence
on the gauge-fixing parameter $a$,\footnote{This parameter labeling different light-cone gauge choices should not be confused with parameter a in the fundamental representation.}
and it is written as\footnote{It
is sufficient to compare only the $\su(2|2)$ invariant
S-matrices.}
$$
{\mathbb S}^{\rm KMRZ}={\mathbb I}+\frac{2\pi
i}{\sqrt{\lambda}}{\mathbb T}\, ,
$$
where ${\mathbb T}$ is a $16\times 16$-matrix (see appendix \ref{appYB}).
It depends on ten non-trivial coefficients ${\rm A },\cdots , {\rm
L}$ which are functions of the momenta $p_1$ and $p_2$. According
to \cite{KMRZ}, these coefficients read as \footnote{The
coefficient ${\rm L}$ presented here coincides with the one in the revised version of
\cite{KMRZ}. } \bea \nonumber
{\rm
A}&=&\frac{1}{4}\Big[(1-2a)(\epsilon(p_2)p_1-\epsilon(p_1)p_2)+\frac{(p_1-p_2)^2}{\epsilon(p_2)p_1-\epsilon(p_1)p_2}\Big]\,
,
\\
\nonumber {\rm B}&=&-{\rm E}=\frac{p_1p_2}{\epsilon(p_2)p_1-\epsilon(p_1)p_2}\, ,\\
\nonumber {\rm C}&=&{\rm
F}=\frac{1}{2}\frac{\sqrt{(\epsilon(p_1)+1)(\epsilon(p_2)+1)}(\epsilon(p_2)p_1-\epsilon(p_1)p_2+p_2-p_1)}
{\epsilon(p_2)p_1-\epsilon(p_1)p_2}\, ,
\\
\nonumber {\rm D}&=&
\frac{1}{4}\Big[(1-2a)(\epsilon(p_2)p_1-\epsilon(p_1)p_2)-\frac{(p_1-p_2)^2}{\epsilon(p_2)p_1-\epsilon(p_1)p_2}\Big]\,
,
\nonumber\\
\nonumber {\rm G}&=&
\frac{1}{4}\Big[(1-2a)(\epsilon(p_2)p_1-\epsilon(p_1)p_2)-\frac{p_1^2-p_2^2}{\epsilon(p_2)p_1-\epsilon(p_1)p_2}\Big]\,
,
\nonumber\\
\nonumber {\rm H}&=&{\rm K}
=\frac{1}{2}\frac{p_1p_2}{\epsilon(p_2)p_1-\epsilon(p_1)p_2}
\frac{(\epsilon(p_1)+1)(\epsilon(p_2)+1)-p_1p_2}{\sqrt{(\epsilon(p_1)+1)(\epsilon(p_2)+1)}}
\, ,
\\
 \nonumber {\rm L}&=&
\frac{1}{4}\Big[(1-2a)(\epsilon(p_2)p_1-\epsilon(p_1)p_2)+\frac{p_1^2-p_2^2}{\epsilon(p_2)p_1-\epsilon(p_1)p_2}\Big]\,
. \nonumber
 \eea
Here $\epsilon(p)=\sqrt{1+p^2}$ is the relativistic energy. To
make a comparison of our string S-matrix  $\S(p_1,p_2)$ with that of \cite{KMRZ} we recall that the string theory choice of the parameters $\eta_{1,2}$ and $\tilde{\eta}_{1,2}$ in (\ref{Smatrix}) is
$$
\eta_1=\eta(p_1)e^{\frac{i}{2}p_2}\, , ~~~~~ \eta_2=\eta(p_2)\, ,
~~~~ \tilde{\eta}_1=\eta(p_1)\, ,
~~~~\tilde{\eta}_2=\eta(p_2)e^{\frac{i}{2}p_1}\, ,
$$
where $\eta(p)=\sqrt{ix^-(p)-ix^+(p)}$. Then, we also have to multiply the S-matrix in (\ref{Smatrix}) by the scalar factor (\ref{scf}) which encodes the dynamical information and the gauge dependence on the parameter $a$. This gives us the string S-matrix we should compare with ${\mathbb S}^{\rm KMRZ}$
\bea\la{cS}
\S_{12}(p_1,p_2) = S_{0}(p_1,p_2)S_{12}(p_1,p_2)\,.
\eea
The near plane-wave expansion of the string S-matrix is constructed by
rescaling the momenta as $p\to \frac{2\pi}{\sqrt{\lambda}}p$ and
keeping in the expansion of $\S(p_1,p_2)$ around $\lambda=\infty$
the first two leading terms.

\medskip
One can check that in the  limit
$\lambda\to\infty$ the matrix $\S(p_1,p_2)$ tends to the
graded identity in accord with the discussion in section 2.  On the other hand, ${\mathbb S}^{\rm KMRZ}$ goes to the identity in this limit. The reason for this discrepancy is in the different conventions used in \cite{KMRZ} and this paper.
Analyzing the definitions, one can see that the S-matrices should be related as follows
$$
{\mathbb S}^{\rm KMRZ}(p_1,p_2)={\cal
P}P\Big(\S(p_1,p_2)\Big)^{-1}|_{{\cal
O}\big(\frac{1}{\sqrt{\lambda}}\big)}\,,
$$
where $P$ and ${\cal P}$ are the permutation and graded
permutation matrices respectively; the product ${\cal P}P$ is the
graded identity.
We then find the perfect agreement between these S-matrices.

In \cite{KMRZ} the S-matrix ${\mathbb S}^{\rm KMRZ}(p_1,p_2)$ was
compared to the near plane-wave expansion of the S-matrix found by
Beisert \cite{B}. Upon a proper choice of the
dressing factor the S-matrices appear to agree up to  terms linear
in momenta; this difference was attributed to the difference in
the definition of the spin chain and string lengths arising from
world-sheet excitations. In our present discussion we never
refered to the gauge theory and we found a complete agreement
between the S-matrix derived from symmetry principles and the one
computed from the near plane-wave string Lagrangian. We note that
it is the proper choice of $\eta$'s in $S(p_1,p_2)$ which is
ultimately responsible for this agreement.

We conclude this section by mentioning the relation of the spin chain
S-matrix $S^{\rm chain}(p_1,p_2)$ (see appendix \ref{appsmat}) with that of Beisert
\cite{B}, $S^{B}(p_1,p_2)$. Choosing
$\tilde{\eta}_1=\eta_1$ and $\tilde{\eta}_2=\eta_2$ in (\ref{Smatrix}), the relation
is as follows
$$
P{\cal P}S_{12}^{\rm chain}(p_2,p_1){\cal P}=S^B_{12}(p_1,p_2)\, .
$$

%%%%%%%%%%%%%%%%%%%%%%%%%%%%%%%%%%%%%%%%%%%%%%%%%%%%%%%%%
\section*{Acknowledgements}
We are grateful to Niklas Beisert, Jan Plefka, Fabian Spill,
Matthias Staudacher and Alessandro Torrielli for valuable comments on the manuscript. The work
of G.~A. was supported in part by the RFBI grant N05-01-00758, by
the grant NSh-672.2006.1, by NWO grant 047017015 and by the INTAS
contract 03-51-6346. The work of S.F.  was supported in part by the Science Foundation
Ireland under Grant No. 07/RFP/PHYF104.
The work of G.~A. and S.~F.~was supported in
part by the EU-RTN network {\it Constituents, Fundamental Forces
and Symmetries of the Universe} (MRTN-CT-2004-005104). The work of
 M.~Z. was supported in part by the grant {\it
Superstring Theory} (MRTN-CT-2004-512194). M.~Z. would like to
thank the Spinoza Insitute for the hospitality during the last
phase of this project.

\section{Appendix}\la{app}

\subsection{Notations}\la{notations}

Throughout the paper we use the standard notations.  The indices
$M$ and ${\dot M}$ run from 1 to 4, and the index $i$ is a
collective index to denote a set $\{ M,{\dot M}\}$. Then we
introduce rows $e^M\,,\ e^{\dM}\,,\ E^i$ and columns $e_M\,,\
e_{\dM}\,,\ E_i$ as \bea
e^M&=&(0,...,1_M,...,0)\,,\quad e_M = (e^M)^\dagger\,,\quad e^M\cdot e_N = \delta^M_N\\
E^i &\equiv& E^{M\dM}=e^M\otimes e^{\dM}\,,\quad E_i \equiv
E_{M\dM}=e_M\otimes e_{\dM}\,, \eea and matrix  unities \bea
e^N_M= e_M\otimes e^N\,,\quad E^i_k = E_k\otimes E^i\,. \eea Note
that $E^k E^i_j = \delta^k_j E^i\,,\ E^i_j E_k = \delta_k^i
E_j\,,\ E^i_j E^k_m = \delta^i_m E^k_j$. They are used to form
columns and rows from the creation and annihilation operators, and
matrices from the S-matrix, and symmetry algebra structure
constants, e.g. \bea A^\dagger= \sum_i A^\dagger_i(p)\,
E^i\,,\quad A = \sum_i A^i(p)\, E_i \,,\quad S_{12}= \sum_{ijkl}
S_{ij}^{kl}(p_1,p_2)\, E^i_k \otimes E^j_l\,.\eea We also define
the following matrix delta-function \bea
 \delta_{12}= \delta(p_1-p_2) \sum_i E_i\otimes E^i\,,
\eea
and use the following convention \bea
A^\dagger_1A^\dagger_2= \sum_{i,j}
A^\dagger_i(p_1)A^\dagger_j(p_2)\,  E^i \otimes E^j\,,\quad
A^\dagger_2A^\dagger_1= \sum_{i,j}
A^\dagger_j(p_2)A^\dagger_i(p_1)\,  E^i \otimes E^j\,. \eea
Finally, if $A,B,C$ are either columns or rows with operator
entries then in the notation $A_1 B_2C_3$ the subscripts $1,2,3$
refer to the location of the columns and row, e.g. if $A =
A^i(p_3)\, E_i\,, \ B=B_i(p_1)E^i\,,\ C=C_i(p_2)E^i$, then
$A_1 B_3 C_2=A^i(p_3)B_k(p_1)C_j(p_2)E_i\otimes E^j\otimes E^k$.

%%%%%%%%%%%%%%%%%%%%%%%%%%
\subsection{S-matrix}\la{appsmat}
Up to a scalar factor the invariant S-matrix satisfying
eqs.(\ref{incondb}), (\ref{incondf}) can be written in the
following form

{\footnotesize
 \bea
 \nonumber
 S(p_1,p_2)&=&\frac{x^-_2-x^+_1}{x^+_2-x^-_1}\frac{\eta_1\eta_2}{\tilde{\eta}_1\tilde{\eta}_2}
\Big(E_{1}^{1}\otimes E_{1}^{1}+E_{2}^{2}\otimes
E_{2}^{2}+E_{1}^{1}\otimes
E_{2}^{2}+E_{2}^{2}\otimes E_{1}^{1}\Big)\\
\nonumber &+&\frac{(x_1^--x_1^+)(x_2^- -x_2^+)(x_2^-+x_1^+)}
{(x^-_1-x^+_2)(x_1^-x_2^--x_1^+x_2^+)}\frac{\eta_1\eta_2}{\tilde{\eta}_1\tilde{\eta}_2}\Big(E_{1}^{1}\otimes
E_{2}^{2}+E_{2}^{2}\otimes E_{1}^{1}-E_{1}^{2}\otimes
E_{2}^{1}-E_{2}^{1}\otimes E_{1}^{2}\Big)
\\
\nonumber &-& \Big(E_{3}^{3}\otimes
E_{3}^{3}+E_{4}^{4}\otimes E_{4}^{4}+E_{3}^{3}\otimes E_{4}^{4}+E_{4}^{4}\otimes E_{3}^{3}\Big)\nonumber\\
\nonumber & +&\frac{(x_1^--x_1^+)(x_2^- -x_2^+)(x_1^-+x_2^+)}
{(x^-_1-x^+_2)(x_1^-x_2^--x_1^+x_2^+)}\Big(E_{3}^{3}\otimes
E_{4}^{4}
+E_{4}^{4}\otimes E_{3}^{3}-E_{3}^{4}\otimes E_{4}^{3}-E_{4}^{3}\otimes E_{3}^{4}\Big)\\
\nonumber &+&\frac{x_2^--x_1^-}{x_2^+
-x_1^-}\frac{\eta_1}{\tilde{\eta}_1}\Big(E_{1}^{1}\otimes
E_{3}^{3}+E_{1}^{1}\otimes E_{4}^{4}+E_{2}^{2}\otimes
E_{3}^{3}+E_{2}^{2}\otimes
E_{4}^{4}\Big)\\
\nonumber &
+&\frac{x_1^+-x_2^+}{x_1^--x_2^+}\frac{\eta_2}{\tilde{\eta}_2}\Big(E_{3}^{3}\otimes
E_{1}^{1}+E_{4}^{4}\otimes
E_{1}^{1}+E_{3}^{3}\otimes E_{2}^{2}+E_{4}^{4}\otimes E_{2}^{2}\Big)\\
\nonumber & +&i\frac{(x_1^--x_1^+)(x_2^--x_2^+)(x_1^+-x_2^+)
}{(x_1^--x_2^+)(1-x_1^-x_2^-)\tilde{\eta}_1\tilde{\eta}_2}
\Big(E_{1}^{4}\otimes E_{2}^{3}+E_{2}^{3}\otimes E_{1}^{4}-E_{2}^{4}\otimes E_{1}^{3}-E_{1}^{3}\otimes E_{2}^{4}\Big)\\
\nonumber
&+&i\frac{x_1^-x_2^-(x_1^+-x_2^+)\eta_1\eta_2}{x_1^+x_2^+(x_1^--x_2^+)(1-x_1^-x_2^-)}
\Big(E_{3}^{2}\otimes E_{4}^{1}+E_{4}^{1}\otimes
E_{3}^{2}-E_{4}^{2}\otimes E_{3}^{1}-E_{3}^{1}\otimes
E_{4}^{2}\Big)\\ \nonumber &+&
\frac{x_1^+-x_1^-}{x_1^--x_2^+}\frac{\eta_2}{\tilde{\eta}_1}\Big(E_{1}^{3}\otimes
E_{3}^{1}+E_{1}^{4}\otimes E_{4}^{1}+E_{2}^{3}\otimes
E_{3}^{2}+E_{2}^{4}\otimes
E_{4}^{2} \Big) \\
\label{Smatrix} &+&\frac{x_2^+
-x_2^-}{x_1^--x_2^+}\frac{\eta_1}{\tilde{\eta}_2}
\Big(E_{3}^{1}\otimes E_{1}^{3}+E_{4}^{1}\otimes
E_{1}^{4}+E_{3}^{2}\otimes E_{2}^{3}+E_{4}^{2}\otimes E_{2}^{4}
\Big)
 \eea
} Here $\eta_i$ depend on the momenta $p_i$ and the parameters
$\xi_i$, and for unitary representations they are given by $ \eta
= \sqrt{i x^- - i x^+} e^{i(\xi+ \vp )}$. In particular, for the
string symmetric choice all $\vp_i=0$, and $\eta_i$ are equal to
\bea\nonumber\mbox{\sc String\ basis:\ }\quad
\eta_1=\eta(p_1)e^{\frac{i}{2}p_2}\, , ~~~ \eta_2=\eta(p_2)\, , ~~
\tilde{\eta}_1=\eta(p_1)\, ,
~~\tilde{\eta}_2=\eta(p_2)e^{\frac{i}{2}p_1}\, , ~~~~~~~~~\eea where
$\eta(p)=\sqrt{ix^-(p)-ix^+(p)}$. For the spin chain choice we
have \bea\nonumber\mbox{\sc Spin\ chain\ basis:\
}\quad\eta_1=\eta(p_1)\, , ~~~ \eta_2=\eta(p_2)\, , ~~
\tilde{\eta}_1=\eta(p_1)\, , ~~\tilde{\eta}_2=\eta(p_2)\,.~~~~~~~~~~\eea The
string theory, $S^{{\rm string}}$, and spin chain, $S^{{\rm
chain}}$, S-matrices are related as  follows \bea\la{tS32}
S_{12}^{{\rm chain}}(p_1,p_2)= U_2(p_1) S_{12}^{{\rm
string}}(p_1,p_2) U_1^\dagger(p_2)\,,\quad U(p)= {\mbox {diag}} (
e^{{i\ov 2}p}, e^{{i\ov 2}p}, 1,1)\,. \eea Since the string theory
S-matrix satisfies the usual YB equation,  the spin chain S-matrix
obeys the twisted YB equation with the twist determined by $U(p)$.

\medskip

Let us now summarize the characteristic properties of the string
S-matrix. Substituting in the supersymmetry generators
$J(p,\zeta,\eta)$ the corresponding values of $\zeta$ and $\eta$
from the string basis we see that the invariance condition for the
string S-matrix takes the form
$$
S(p_1,p_2)\Big(J(p_1)\otimes e^{\frac{i}{2}p_2}+\Sigma\otimes
J(p_2) \Big)=\Big(J(p_1)\otimes \Sigma+e^{\frac{i}{2}p_1}\otimes
J(p_2) \Big)S(p_1,p_2)\, ,
$$
where $J(p)$ are the supersymmetry generators $\{Q_{\a}^a\}$ with
$\xi=\varphi=0$, cf. eq.(\ref{urep}). From the explicit formula
for the S-matrix we see that it depends not only on $x^{\pm}(p)$
but also on the exponential factors $e^{\frac{i}{2}p}$. While
$x^{\pm}$ are periodic functions of $p$ with the period $2\pi$ the
exponential factors are not. As the result, the string S-matrix is
not periodic under the shift $p\to p+2\pi$, rather it exhibits the
following monodromy properties \bea\nonumber
S_{12}(p_1,p_2+2\pi)&=&-S_{12}(p_1,p_2)\Sigma_1\, , \\
S_{12}(p_1+2\pi,p_2)&=&-\Sigma_2S_{12}(p_1,p_2)\, , \nonumber\eea
where $\Sigma_1=\Sigma\otimes {\mathbb I}$ and $\Sigma_2={\mathbb
I}\otimes \Sigma$. The compatibility of these equations with the
invariance condition is guaranteed by the fact that all
supersymmetry generators $J$ obey the relation $\{\Sigma,J\}=0$.
Finally, we note that under the shift by $2\pi$ any of the three
momenta $p_1,p_2,p_3$ entering the Yang-Baxter equation, this
equation turns into itself because the S-matrix enjoys the
following special property
$$
[S_{12}(p_1,p_2),\Sigma\otimes\Sigma]=0\, .
$$

\medskip

To describe the $\AdS$ light-cone string theory the S-matrix
(\ref{Smatrix}) has to be multiplied by the scalar factor,
$S_0(p_1,p_2)$, of the following form \cite{AF06} \bea \la{scf}
S_0(p_1,p_2)^2 = \frac{x^+_2-x^-_1}{x^+_1-x^-_2}\, {1-{1\ov x_1^+
x_2^-}\ov 1-{1\ov x_1^- x_2^+}}\, e^{i\theta(p_1,p_2)}\, e^{i
a(p_1  \epsilon_2-p_2 \epsilon_1)}\,. \eea Here the
gauge-independent dressing phase $\theta(p_1,p_2)$ \cite{AFS} is a
two-form on the vector space of conserved charges $q_n(p)$ \bea
\nonumber \theta(p_1,p_2) = \sum_{r=2}^\infty \sum_{n=0}^\infty
c_{r,r+1+2n}(\l)\left( q_r(p_1)q_{r+1+2n}(p_2) -
q_r(p_2)q_{r+1+2n}(p_1) \right)\, , \eea
and $a$ is the parameter
of the generalized light-cone gauge \cite{magnon}, and $\epsilon_i
= H(p_i)$ is the energy of a one-particle state. The
gauge-dependent factor solves the homogeneous crossing equation
\cite{AF06}.

\medskip

The canonical $\su(2|2)^2$ invariant S-matrix is the following tensor product
\bea\la{sfull}
{\cal S} (p_1,p_2) = S_0(p_1,p_2)^2\, S(p_1,p_2)\otimes S(p_1,p_2)\,.
\eea
In the limit $p_1=p_2$ the matrix $S_{12}(p_1,p_2)$ turns into
$-P_{12}$, where $P_{12}$ is the permutation, and the scalar factor $S_0(p_1,p_2)^2$ becomes $-1$. Thus, the full S-matrix
becomes in this limit the minus permutation.

\subsection{Symmetries from the Yang-Baxter equation}
In this section we explain how the knowledge of the S-matrix
satisfying the Yang-Baxter equation allows one to reconstruct the
corresponding representation of the symmetry algebra which, in the
present case, is the centrally extended $\su(2|2)_{H,C}$. In our
exposition we closely follow  the general discussion due to
Bernard and Leclair \cite{Bernard:1990tt} which is also aimed to
reveal the higher non-local symmetries of the S-matrix.

\medskip

Let us introduce an associative algebra generated by the symbols
${\rm T}_{ij}(p,\mu)$, where $i,j=1,\ldots, 4$, modulo the
following relations \bea\label{STT} S_{12}(p_1,p_2){\rm
T}_1(p_1,\mu){\rm T}_2(p_2,\mu)={\rm T}_2(p_2,\mu){\rm
T}_1(p_1,\mu)S(p_1,p_2)\, . \eea Here the matrix spaces 1 and 2
are considered as auxiliary ones and we use the notation ${\rm
T}_1={\rm T}\otimes {\mathbb I}$ and ${\rm T}_2={\mathbb I}\otimes
{\rm T}$. The variable $\mu$ is a (spectral) parameter of the
representation. The absence of cubic and higher order relations
between the algebra generators is guaranteed by the Yang-Baxter
equation for the S-matrix. A particular 4-dimensional
representation of this algebra is provided by the S-matrix itself:
$$
{\rm T}_1(p_1,\mu)=S_{13}(p_1,p_3)\, ~~~~~{\rm with}~~~~~~p_3=\mu\, .
$$
Upon expanding around a special point in the spectral parameter
plane the algebra (\ref{STT}) is expected to produce the
generators of both the local and non-local symmetries of the
theory. In our model the only distinguished point corresponds to
$p_3=0$.
%Indeed, the FZ oscillators evaluated at $p=0$ correspond
%to zero-modes of the world-sheet fields, i.e. they are, in fact,
%represent the supergravity modes.
Thus, we have to understand the
expansion of the Yang-Baxter equation around $p_3=0$.

\medskip

Since the relation (\ref{xpm}) is quadratic in $x^{+}$ and
$x^{-}$, for $x^{\pm}$ as functions of $p$ we find two different
solutions. In the limit $p\to 0$ the first solution expands as
\bea \la{xpm1} x^{\pm}=\frac{1}{gp}\pm
\frac{i}{2g}+\Big(g-\frac{1}{12g}\Big)p+\cdots\, , \eea while the
second one yields \bea\la{xmp2} x^{\pm}=-g p  \mp
\frac{ig}{2}p^2+\cdots \eea In what
 follows we pick up the first solution because, in opposite to the second one, this solution implies that in
 the limit $p\to\ 0$ the variable $\eta= \frac{1}{\sqrt{g}}$ is real
 for a positive value of the coupling constant $g$, i.e. that the
 corresponding representation of $\su(2|2)$ is unitary. Thus,
 taking into account eq.(\ref{xpm1}) the canonical $\su(2|2)$-S-matrix
simplifies in the limit $p_2\to 0$ to \bea \nonumber
S_{12}(p,0)=e^{-\frac{i}{2}p}~{\mathbb I}\otimes
 (E_{1}^{1}+E_{2}^{2})+\Sigma\otimes(E_{3}^{3}+E_{4}^{4})\, .
\eea
%This diagonal S-matrix describes the scattering between the
%string and the supergravity modes.
If we put in the last formula
$p=0$ then the corresponding S-matrix turns into the graded unity.
It is interesting to note that if we would take the limit
$p_1=p_2=p$ first and then $p=0$ we would obtain  a different
result which is the minus permutation. Thus, the order in which
the limits are taken matters.

\medskip

Consider now the Yang-Baxter equation
$$
S_{12}(p_1,p_2)S_{13}(p_1,p_3)S_{23}(p_2,p_3)=S_{23}(p_2,p_3)S_{13}(p_1,p_3)S_{12}(p_1,p_2)\,
$$
and expand it in power series around the special point $p_3=0$.
The leading term of this expansion produces an equation
$$
S_{12}(p_1,p_2)S_{13}(p_1,0)S_{23}(p_2,0)=S_{23}(p_2,0)S_{13}(p_1,0)S_{12}(p_1,p_2)\,
$$
which is obviously satisfied because it is the Yang-Baxter
equation considered for the fixed value of $p_3$. Next, we take
the subleading equation, multiply its both sides from the left by
$S_{13}^{-1}S_{23}^{-1}$ and make use of the Yang-Baxter
equation. This gives the following relation \bea\la{YBDif}
S_{12}\Big(S_{23}^{-1}S_{13}^{-1}\pa_3
S_{13}S_{23}+S_{23}^{-1}\pa_3 S_{23}\Big)=\Big(S_{13}^{-1}\pa_3
S_{13}+S_{13}^{-1}S_{23}^{-1}\pa_3 S_{23}S_{13}\Big)S_{12}\, ,
\eea where we used the shorthand notation
$$
\pa_3 S_{13}\equiv \frac{\pa S_{13}(p_1,p_3)}{\pa p_3}|_{p_3=0}\,
$$
and analogously for $\pa_3 S_{23}$.  Equation (\ref{YBDif})
involves three matrix spaces. We project out the third space by
multiplying both sides with a constant matrix $\Upsilon$ living in
the third space and take the matrix trace over the third space. We
get \bea\la{kinrestored} && S_{12}\Big[{\rm Tr
}_3\big(S_{23}^{-1}S_{13}^{-1}\pa_3
S_{13}S_{23}\Upsilon_3\big)+{\mathbb I}\otimes {\rm Tr
}_3\big(S_{23}^{-1}\pa_3
S_{23}\Upsilon_3\big)\Big]=\\
\nonumber &&~~~~~~~~~~~~~~~~~~~=\Big[{\rm Tr
}_3\big(S_{13}^{-1}\pa_3 S_{13}\Upsilon_3\big)\otimes {\mathbb I}
+{\rm Tr}_3\big(S_{13}^{-1}S_{23}^{-1}\pa_3
S_{23}S_{13}\Upsilon_3\big)\Big]S_{12}\, . \eea The derivative of
the S-matrix entering the last equation can be easily found if one
takes into account the formulae \bea \frac{\pa x^{\pm}}{\pa
p~}=\frac{ix^{\pm 2}(1-x^{\pm 2 })}{(x^+-x^-)(1+x^{+}x^{-})} \, ,
~~~~~~~~~~ \frac{\pa \eta }{\pa
p}=\frac{1}{2\eta}\frac{x^++x^-}{1+x^+x^-}\, ,\eea where
$\eta=\sqrt{ix^--ix^+}$. Evaluating (\ref{kinrestored}) one finds
that this equation is equivalent to the invariance condition
(\ref{incondb}) for the set of bosonic generators provided one
makes a proper choice for the matrix $\Upsilon$: \bea\nonumber
&&L_a^b\leftrightarrow i E_a^b\, ,~~~~a\neq b,~~~a,b=1,2;~~~~~~
R_{\a}^{\b}\leftrightarrow -i E_\a^\b\, ,~~~~\a\neq \b,~~~\a,\b=3,4;\\
\nonumber &&L_1^1=-L_2^2\leftrightarrow\frac{i}{2}(E_1^1-E_2^2)\,
,~~~~~~~~~~~~~~~
R_1^1=-R_2^2\leftrightarrow-\frac{i}{2}(E_3^3-E_4^4)\, .\eea Also,
the Hamiltonian $H$ is generated by $\Upsilon=i\Sigma$.

\medskip

Now we would like to rederive the invarince condition
(\ref{incondf}) involving the supersymmetry generators. To this
end we recall that $S_{12}$ commutes with $\Sigma_1\Sigma_2$ so
that we can rewrite equation (\ref{YBDif}) in the form
\bea\nonumber && S_{12}\Big(S_{23}^{-1}S_{13}^{-1}\pa_3
S_{13}S_{23}\Sigma_1\Sigma_2+\Sigma_1S_{23}^{-1}\pa_3
S_{23}\Sigma_2\Big)=\\
\nonumber
&&~~~~~~~~~~~~~~~~~~~~~~~~~~~~~~~~~~~~~~~~~~~=\Big(S_{13}^{-1}\pa_3
S_{13}\Sigma_1\Sigma_2+S_{13}^{-1}S_{23}^{-1}\pa_3
S_{23}S_{13}\Sigma_1\Sigma_2\Big)S_{12}\, . \eea  Again we project
out the third space in the last equation by multiplying both sides
with a constant matrix $\Upsilon_3$ and taking the trace over the
third space: \bea\nonumber && S_{12}\Big[{\rm Tr
}_3\big(S_{23}^{-1}S_{13}^{-1}\pa_3
S_{13}S_{23}\Upsilon_3\big)\Sigma_1\Sigma_2+\Sigma\otimes {\rm Tr
}_3\big(S_{23}^{-1}\pa_3
S_{23}\Upsilon_3\big)\Sigma_2\Big]=\\
\nonumber &&~~~~~~~~~~~~~~~~~~~=\Big[{\rm Tr
}_3\big(S_{13}^{-1}\pa_3 S_{13}\Upsilon_3\big)\Sigma_1\otimes
\Sigma+{\rm Tr}_3\big(S_{13}^{-1}S_{23}^{-1}\pa_3
S_{23}S_{13}\Upsilon_3\big)\Sigma_1\Sigma_2\Big]S_{12}\, . \eea
One can see that the last equation indeed reduces to \bea
S_{12}\big(Q_{\a}^a(p_1)\otimes e^{\frac{i}{2}p_2}+\Sigma\otimes
Q_{\a}^a(p_2) \big)&=&\big(Q_{\a}^a(p_1)\otimes
\Sigma+e^{\frac{i}{2}p_1}\otimes Q_{\a}^a(p_2) \big)S_{12}\,
\nonumber
%\\
%S_{12}\big(Q_{a}^{\dagger\a}(p_1)\otimes
%e^{-\frac{i}{2}p_2}+\Sigma\otimes Q_{a}^{\dagger\a}(p_2)
%\big)&=&\big(Q_{a}^{\dagger\a}(p_1)\otimes
%\Sigma+e^{-\frac{i}{2}p_1}\otimes Q_{a}^{\dagger\a}(p_2)
%\big)S_{12}\, ,\nonumber
\eea provided one makes the following choice of $\Upsilon$:
$Q_{\a}^a\leftrightarrow -i E_{\a}^a$, where $a=1,2$ and $\a=3,4$.
Analogously, $Q_{a}^{\dagger\a}\leftrightarrow i E_{a}^{\a }$.
%, where $a=3,4$ and $\a=1,2$.
Thus, we have shown that the symmetry
algebra $\su(2|2)_{H,C}$ (more precisely, its fundamental
representation) arises upon expansion of the S-matrix around the
special point $p=0$.

\medskip

One can expand the Yang-Baxter equation further and find the
fundamental representation for the higher (non-local) symmetries
commuting with the S-matrix. We will not proceed with this
expansion because, as it is clear from the our construction, the
form of the S-matrix is fixed uniquely up to a scalar phase from
the requirement of the $\su(2|2)_{H,C}$-symmetry, and the additional nonlocal symmetries (in the fundamental representation) do not seem to lead to any additional equations for the scalar phase.

\subsection{Hopf algebra interpretation}\la{hopf}
In this appendix we discuss a Hopf algebra interpretation
\footnote{We would like to thank J. Plefka and F. Spill for an
interesting discussion of the subject.} of the S-matrix invariance
condition (\ref{incondf}). The form of the comultiplication
depends on the choice of the basis of the fundamental
representation, and we restrict our attention to the symmetric
string theory basis.

Fist of all, we stress that to define a nontrivial co-product one
should consider not the centrally-extended super-Lie algebra
$\su(2|2)_{H,C}$ but a unitary graded associative algebra $\cA$
generated by the even rotation generators $\bL_a{}^b\,,\ \bR_\a{}^\b$,
the odd supersymmetry generators $\bQ_\a{}^a,\,\ \bQ_a^{\dagger}{}^\a$
and two central elements $\bH$ and $\hP$ subject to the algebra
relations (\ref{su22}) with the central elements $\bC$ and
$\bC^\dagger$ expressed through the world-sheet momentum $\hP$ as
follows \bea \label{Cc1} \bC=ig\,(e^{i\hP}-{\rm id})\,,\quad
\bC^\dagger=-ig\,(e^{-i\hP}-{\rm id})\, . \eea Then the
comultiplication can be defined on the generators as follows \bea
\nonumber \Delta({\bf J}) &=&{\bf J}\otimes {\rm id} + {\rm id}\otimes
{\bf J}\, \quad {\rm for\ any\ even\ generator}\,,\\\la{coprod}
\Delta(\bQ_\a{}^a) &=&\bQ_\a{}^a\otimes e^{i\hP/2} + {\rm id}\otimes
\bQ_\a{}^a\,,\\\nonumber \Delta(\bQ_a^{\dagger}{}^\a)
&=&\bQ_a^{\dagger}{}^\a\otimes e^{-i\hP/2} + {\rm id}\otimes
\bQ_a^{\dagger}{}^\a\,. \eea Here we use the graded tensor product,
that is for any algebra elements $a,b,c,d$ $$(a\otimes b)(c\otimes d)
= (-1)^{\epsilon(b)\epsilon(c)}(ac\otimes bd),$$ where $\epsilon(a)=0$
if $a$ is an even element, and $\epsilon(b)=-1$ if $a$ is an odd
element of the algebra $\cA$. These comultiplication rules are
equivalent (up to a twist and some redefinitions of the supersymmetry
generators and the central elements $\bC$ and $\bC^\dagger$) to the
ones discussed in \cite{Gomez:2006va}.  In addition to the co-product, we
also use the standard definitions of unit and co-unit.

\medskip

Let us show that the comultiplication  agrees with the form of the
two-particle structure constants appearing in (\ref{incondf}). Let
$V$ be the fundamental representation of the algebra $\cA$.
Obviously, it is isomorphic to the representation $V(p,1)$ of
$\su(2|2)_{H,C}$. The vector space $V$ has a natural grading. The
action of, say, the supersymmetry generators $\bQ_\a{}^a$ on the
tensor product $V\otimes V$ is given by the comultiplication
(\ref{coprod}) \bea \Delta(\bQ_\a{}^a)\cdot e_M\otimes e_N &=&
(\bQ_\a{}^a\otimes e^{i\hP/2} +  {\rm id}\otimes \bQ_\a{}^a)\cdot
e_M\otimes e_N
\\\nonumber &=&  \bQ_\a{}^a\cdot e_M\otimes e^{i\hP/2}\cdot e_N +
\Sigma\cdot e_M\otimes \bQ_\a{}^a\cdot e_N\,. \eea Now one can
recognize that the two-particle representation coincides with the
one appearing on the l.h.s. of (\ref{incondf}).

The antipode can be easily found by using (\ref{coprod}). For any
even generator including $\bH$ and $\hP$ it acts as
$$
S(\bJ)=-{\bf J}\, ,
$$
while on supersymmetry generators it is defined as
$$ S(\bQ_\a{}^a) =-\bQ_\a{}^a\, e^{-i\hP/2}\,,\quad
S(\bQ_a^{\dagger}{}^\a) =-\bQ_a^{\dagger}{}^\a\, e^{i\hP/2} \,. $$
Finally, let us mention that the ZF algebra does not have a Hopf
algebra structure.

\subsection{The twisted Yang-Baxter equation}\la{appYB}
In this appendix we derive the most general form of the twisted YB equation.

Every fundamental representation $V$ is parametrized by momentum $p$ and the
label $\zeta$ (and $\eta$ which will not be always  shown explicitly in the consideration below) which may be
$p$-dependent. The invariant $S$-matrix is defined as
$$
S_{12}(p_1,\zeta_1;p_2,\zeta_2): ~~~~~ V(p_1,\zeta_1)\otimes
V(p_2,\zeta_2)\to V(p_1,\tzeta_{1})\otimes V(p_2,\tzeta_2)\,,
$$
where the vector spaces $V$ are considered as columns.
Then, we must require the two tensor product spaces $V(p_1,\zeta_1)\otimes
V(p_2,\zeta_2)$ and $V(p_1,\tzeta_{1})\otimes V(p_2,\tzeta_2)$ be isomorphic to each other and carry the same representation of the centrally extended $\su(2|2)$. This implies that both spaces must have the same values of the central charges and satisfy the
shortening conditions \cite{Bn}. Then, one finds
\bea
\label{z1} \tzeta_1&=&e^{ip_2}\zeta_2
\frac{(1-e^{ip_1})\zeta_1+(1-e^{ip_2})\zeta_2}{e^{ip_1}\zeta_1+e^{ip_2}\zeta_2-e^{i(p_1+p_2)}(\zeta_1+\zeta_2)}
\\
\label{z2}\tzeta_2&=&e^{ip_1}\zeta_1
\frac{(1-e^{ip_1})\zeta_1+(1-e^{ip_2})\zeta_2}{e^{ip_1}\zeta_1+e^{ip_2}\zeta_2-e^{i(p_1+p_2)}(\zeta_1+\zeta_2)}\eea In particular, we see that
$$
\frac{\tzeta_2}{\tzeta_{1}}=\frac{e^{ip_1}\zeta_1}{e^{ip_2}\zeta_2}\,.
$$
The S-matrix satisfies the invariance condition (\ref{incondf}) which in our case takes the following form for bosonic generators
\bea\nonumber
S_{12}\big(J(p_1;\zeta_1,\eta_1) \otimes \mathbb{I} + \mathbb{I} \otimes J(p_2;\zeta_2,\eta_2)\big)=\big( J(p_1;\tzeta_1,\tilde{\eta}_1) \otimes \mathbb{I} + \mathbb{I} \otimes J(p_2;\tzeta_2,\tilde{\eta}_2)\big)  S_{12}\,,
%\\ \la{incondsu22b}
\eea
and for fermionic generators
\bea
\nonumber
S_{12}\big(J(p_1;\zeta_1,\eta_1) \otimes \mathbb{I} + \Sigma \otimes J(p_2;\zeta_2,\eta_2)\big)=\big( J(p_1;\tzeta_1,\tilde{\eta}_1) \otimes \Sigma + \mathbb{I} \otimes J(p_2;\tzeta_2,\tilde{\eta}_2)\big)  S_{12}\,,
%\\\la{incondsu22f}
\eea where $\Sigma = {\rm diag} (1,1,-1,-1)\,$ takes into account
the fermionic statistics. These equations can be easily solved and
lead to a unique (up to a scalar factor) solution. In terms of
standard matrix unities this solution reads as

{\footnotesize
 \bea
 \nonumber
 S(p_1,\zeta_1;p_2,\zeta_2)&=&a_1
\Big(E_{1}^{1}\otimes E_{1}^{1}+E_{2}^{2}\otimes
E_{2}^{2}+E_{1}^{1}\otimes
E_{2}^{2}+E_{2}^{2}\otimes E_{1}^{1}\Big)\\
\nonumber &+&a_2\Big(E_{1}^{1}\otimes E_{2}^{2}+E_{2}^{2}\otimes
E_{1}^{1}-E_{1}^{2}\otimes E_{2}^{1}-E_{2}^{1}\otimes
E_{1}^{2}\Big)
\\
\nonumber &+& a_3\Big(E_{3}^{3}\otimes
E_{3}^{3}+E_{4}^{4}\otimes E_{4}^{4}+E_{3}^{3}\otimes E_{4}^{4}+E_{4}^{4}\otimes E_{3}^{3}\Big)\nonumber\\
\nonumber & +&a_4\Big(E_{3}^{3}\otimes E_{4}^{4}
+E_{4}^{4}\otimes E_{3}^{3}-E_{3}^{4}\otimes E_{4}^{3}-E_{4}^{3}\otimes E_{3}^{4}\Big)\\
\nonumber &+&a_5\Big(E_{1}^{1}\otimes E_{3}^{3}+E_{1}^{1}\otimes
E_{4}^{4}+E_{2}^{2}\otimes E_{3}^{3}+E_{2}^{2}\otimes
E_{4}^{4}\Big)\\
\nonumber & +&a_6\Big(E_{3}^{3}\otimes E_{1}^{1}+E_{4}^{4}\otimes
E_{1}^{1}+E_{3}^{3}\otimes E_{2}^{2}+E_{4}^{4}\otimes E_{2}^{2}\Big)\\
\nonumber & +&a_7
\Big(E_{1}^{4}\otimes E_{2}^{3}+E_{2}^{3}\otimes E_{1}^{4}-E_{2}^{4}\otimes E_{1}^{3}-E_{1}^{3}\otimes E_{2}^{4}\Big)\\
\nonumber &+&a_8 \Big(E_{3}^{2}\otimes E_{4}^{1}+E_{4}^{1}\otimes
E_{3}^{2}-E_{4}^{2}\otimes E_{3}^{1}-E_{3}^{1}\otimes
E_{4}^{2}\Big)\\ \nonumber &+& a_9\Big(E_{1}^{3}\otimes
E_{3}^{1}+E_{1}^{4}\otimes E_{4}^{1}+E_{2}^{3}\otimes
E_{3}^{2}+E_{2}^{4}\otimes
E_{4}^{2} \Big) \\
\label{Smatrixgen} &+&a_{10}\Big(E_{3}^{1}\otimes
E_{1}^{3}+E_{4}^{1}\otimes E_{1}^{4}+E_{3}^{2}\otimes
E_{2}^{3}+E_{4}^{2}\otimes E_{2}^{4} \Big)\, ,
 \eea
} where the coefficients $a_i$ are found to be {\footnotesize\bea
\nonumber
a_1&=&\frac{\big(x_1^+\zeta_1-x_2^+\zeta_2\big)\big(x_2^-x_1^+\zeta_1+x_1^-x_2^+\zeta_2-x_1^-x_2^-(\zeta_1+\zeta_2)\big)\eta_1\eta_2}
{\big(x_2^-\zeta_1-x_1^-\zeta_2\big)\big(x_2^-x_1^+\zeta_1+x_1^-x_2^+\zeta_2-x_1^+x_2^+(\zeta_1+\zeta_2)\big)\tilde{\eta}_1\tilde{\eta}_2}
\, ,\\
\nonumber a_2&=&
\frac{\big(x_1^--x_1^+\big)\big(x_2^--x_2^+\big)\big(x_1^+\zeta_1+x_2^+\zeta_2\big)\big(x_1^+x_2^-\zeta_1^2
-x_1^-x_2^+\zeta_2^2\big)\eta_1\eta_2}{\big(x_2^-\zeta_1-x_1^-\zeta_2\big)
\big(x_2^-x_1^+\zeta_1+x_1^-x_2^+\zeta_2-x_1^+x_2^+(\zeta_1+\zeta_2)\big)^2\tilde{\eta}_1\tilde{\eta}_2}\, ,\\
\nonumber a_3&=&-1\, ,\\
\nonumber a_4&=&
\frac{\big(x_1^+-x_1^-\big)\big(x_2^--x_2^+\big)\big(x_2^-\zeta_1+x_1^-\zeta_2\big)
\big(x_2^-x_1^+\zeta_1^2-x_1^-x_2^+\zeta_2^2\big)}{
\big(x_2^-\zeta_1-x_1^-\zeta_2\big)
\big(x_2^-x_1^+\zeta_1+x_1^-x_2^+\zeta_2-x_1^-x_2^-(\zeta_1+\zeta_2)\big)
\big(x_2^-x_1^+\zeta_1+x_1^-x_2^+\zeta_2-x_1^+x_2^+(\zeta_1+\zeta_2)\big) }\, ,\\
\nonumber a_5&=&-
\frac{\zeta_2\big((x_1^-x_2^-x_1^++x_2^-x_1^+x_2^+-x_1^-x_2^+x_2^--x_1^-x_1^+x_2^+)
\zeta_1+x_1^-x_2^+(x_1^--x_2^--x_1^++x_2^+)\zeta_2 \big)\eta_1}
{\big(x_2^-\zeta_1-x_1^-\zeta_2\big)\big(x_2^-x_1^+\zeta_1+x_1^-x_2^+\zeta_2-x_1^+x_2^+(\zeta_1+\zeta_2)\big)\tilde{\eta}_1}
\, , \\
 \nonumber
a_6&=&-\frac{\zeta_1\big(x_2^-x_1^+(x_1^--x_2^--x_1^++x_2^+)\zeta_1+(x_1^-x_2^-x_1^++x_2^-x_1^+x_2^+
-x_1^-x_2^+x_2^--x_1^-x_1^+x_2^+)\zeta_2\big)\eta_2}
{\big(x_2^-\zeta_1-x_1^-\zeta_2\big)\big(x_2^-x_1^+\zeta_1+x_1^-x_2^+\zeta_2-x_1^+x_2^+(\zeta_1+\zeta_2)\big)\tilde{\eta}_2}\,
,\\
\nonumber
a_7&=&\frac{\big(x_1^--x_1^+\big)\big(x_2^--x_2^+\big)\big(x_1^-x_2^-x_1^++x_2^-x_1^+x_2^+-x_1^-x_2^+x_2^--x_1^-x_1^+x_2^+\big)
\big(x_2^-x_1^+\zeta_1^2-x_1^-x_2^+\zeta_2^2\big)\zeta_1\zeta_2}{x_1^-x_2^-\big(x_2^-\zeta_1-x_1^-\zeta_2\big)
\big(x_2^-x_1^+\zeta_1+x_1^-x_2^+\zeta_2-x_1^+x_2^+(\zeta_1+\zeta_2)\big)^2\tilde{\eta}_1\tilde{\eta}_2}\,
, \nonumber \eea \bea
 \nonumber
a_8&=&-\frac{x_1^-x_2^-\big(x_1^--x_2^--x_1^++x_2^+\big)\big(x_2^-x_1^+\zeta_1^2-x_1^-x_2^+\zeta_2^2\big)\eta_1\eta_2}{
\big(x_2^-\zeta_1-x_1^-\zeta_2\big)
\big(x_2^-x_1^+\zeta_1+x_1^-x_2^+\zeta_2-x_1^-x_2^-(\zeta_1+\zeta_2)\big)
\big(x_2^-x_1^+\zeta_1+x_1^-x_2^+\zeta_2-x_1^+x_2^+(\zeta_1+\zeta_2)\big)
}\, ,
\\
\nonumber
a_9&=&\frac{\big(x_1^+-x_1^-\big)\big(x_2^-x_1^+\zeta_1^2-x_1^-x_2^+\zeta_2^2\big)\eta_2}
{\big(x_2^-\zeta_1-x_1^-\zeta_2\big)\big(x_2^-x_1^+\zeta_1+x_1^-x_2^+\zeta_2-x_1^+x_2^+(\zeta_1+\zeta_2)\big)\tilde{\eta_1}}\,
,
\\
\nonumber
a_{10}&=&\frac{\big(x_2^+-x_2^-\big)\big(x_2^-x_1^+\zeta_1^2-x_1^-x_2^+\zeta_2^2\big)\eta_1}
{\big(x_2^-\zeta_1-x_1^-\zeta_2\big)\big(x_2^-x_1^+\zeta_1+x_1^-x_2^+\zeta_2-x_1^+x_2^+(\zeta_1+\zeta_2)\big)\tilde{\eta_2}}\,
. \eea }
%\medskip
If $\zeta_1=e^{ip_2}\,,\ \zeta_2=1$ then the S-matrix coincides
with (\ref{Smatrix}).
%For generic values of $\zeta_k$ the S-matrix
%(\ref{Smatrixgen}) obeys the following identity \bea\la{smg}
%S_{12}(p_1,\zeta_1;p_2,\zeta_2) = F_{12}(\xi_2)
%S_{12}\Big(p_1,{\zeta_1\ov
%\zeta_2};p_2,1\Big)F_{12}^{-1}(\xi_2)\,, \eea where $\xi_k \equiv
%{1\ov i}\ln\zeta_k$, and $F_{12}$ is the twist operator
%(\ref{twisto}).

\medskip

To derive the twisted YB equation for the S-matrix we start from
the sequence of  three spaces as
$$
V(p_1,\zeta_1)\otimes V(p_2,\zeta_2)\otimes V(p_3,\zeta_3)\,.
$$
We would like to bring it to the form $V(p_1,\ttz_1)\otimes
V(p_2,\ttz_2)\otimes V(p_3,\ttz_3)$ by permuting the spaces  with
the help of the $S$-matrix. This can be done it two different
ways. The first way is to perform the following  three successive
operations
$$ S_{12}(p_1,\zeta_1;p_2,\zeta_2):
~~~V(p_1,\zeta_1)\otimes V(p_2,\zeta_2)\otimes V(p_3,\zeta_3)
\to V(p_1,\tzeta^{{\ell}}_1)\otimes V(p_2,\tzeta^{{\ell}}_2)\otimes V(p_3,\zeta_3)\\
$$
$$ S_{13}(p_1,\tzeta^{{\ell}}_1;p_3,\zeta_3):
~~~V(p_1,\tzeta^{{\ell}}_1)\otimes V(p_2,\tzeta^{{\ell}}_2)\otimes V(p_3,\zeta_3)
\to V(p_1,\ttzeta_1{}^{{\ell}})\otimes V(p_2,\tzeta^{{\ell}}_2)\otimes V(p_3,\tzeta^{{\ell}}_3)\\
$$
$$ S_{23}(p_2,\tzeta^{{\ell}}_2;p_3,\tzeta^{{\ell}}_3):
~~~V(p_1,\ttzeta_1{}^{{\ell}})\otimes
V(p_2,\tzeta^{{\ell}}_2)\otimes V(p_3,\tzeta^{{\ell}}_3)
\to V(p_1,\ttzeta_1{}^{{\ell}})\otimes V(p_2,\ttzeta_2{}^{{\ell}})\otimes V(p_3,\ttzeta_3{}^{{\ell}})\\
$$
Here to find expressions for $\tzeta^{{\ell}}_i$ and
$\ttz_i{}^{{\ell}}$ we need to use the formulas (\ref{z1}) and
(\ref{z2}). Thus we have applied the following scattering operator
$$
S_{23}(p_2,\tzeta^{{\ell}}_2;p_3,\tzeta^{{\ell}}_3)S_{13}(p_1,\tzeta^{{\ell}}_1;p_3,\zeta_3)S_{12}(p_1,\zeta_1;p_2,\zeta_2)
$$
and ended up with the space
$$V(p_1,\ttzeta_1{}^{{\ell}})\otimes V(p_2,\ttzeta_2{}^{{\ell}})\otimes V(p_3,\ttzeta_3{}^{{\ell}})
$$
The second way to achieve the same final space is to perform
another three successive operations
$$ S_{23}(p_2,\zeta_2;p_3,\zeta_3):
~~~V(p_1,\zeta_1)\otimes V(p_2,\zeta_2)\otimes V(p_3,\zeta_3)
\to V(p_1,\zeta_1)\otimes V(p_2,\tzeta^{r}_2)\otimes V(p_3,\tzeta^r_3)\\
$$
$$ S_{13}(p_1,\zeta_1;p_3,\tzeta^r_3):
~~~V(p_1,\zeta_1)\otimes V(p_2,\tzeta^{r}_2)\otimes V(p_3,\tzeta^r_3)
\to V(p_1,\tzeta_1^{r})\otimes V(p_2,\tzeta^{r}_2)\otimes V(p_3,\ttzeta_3{}^{r})\\
$$
$$ S_{12}(p_1,\tzeta^{r}_1;p_2,\tzeta^{r}_2):
~~~ V(p_1,\tzeta_1^{r})\otimes V(p_2,\tzeta^{r}_2)\otimes V(p_3,\ttzeta_3{}^{r})
\to V(p_1,\ttzeta_1{}^{r})\otimes V(p_2,\ttzeta_2{}^{r})\otimes V(p_3,\ttzeta_3{}^{r})\\
$$
Remarkably, $\ttz_i{}^{{\ell}} = \ttz_i{}^{r}$, and therefore in
both  cases we ended up with the same space. Thus, the exact form
of the twisted Yang-Baxter equation is \bea\la{YBT}
&&S_{23}(p_2,\tzeta^{{\ell}}_2;p_3,\tzeta^{{\ell}}_3)S_{13}(p_1,\tzeta^{{\ell}}_1;p_3,\zeta_3)S_{12}(p_1,\zeta_1;p_2,\zeta_2)
=\\\nonumber
&&~~~~~~~~~~~~~~~~~~=S_{12}(p_1,\tzeta^{r}_1;p_2,\tzeta^{r}_2)S_{13}(p_1,\zeta_1;p_3,\tzeta^r_3)S_{23}(p_2,\zeta_2;p_3,\zeta_3)\,.
\eea

\medskip
If we choose the parameters $\zeta_i$ as follows
\bea\nonumber
 \zeta_3 = 1\,,\quad \zeta_2=e^{ip_2}\,,\quad \zeta_1=e^{i(p_2+p_3)}\,,
 \eea
 and the string theory choice of the parameters $\eta_i$ then the twisted YB equation (\ref{YBT}) becomes the usual one. If we choose the spin chain parameters $\eta_i$ then we reproduce the  twisted YB equation (\ref{TYB}).
\medskip

We conclude this appendix by noting that sometimes it is
convenient to visualize the S-matrix as an explicit $16\times 16$
matrix

{\scriptsize
\begin{eqnarray}
 S(p_1,\zeta_1;p_2,\zeta_2) \equiv \left( \begin{array}{ccccccccccccccccccc}
a_1&0&0&0&|&0&0&0&0&|&0&0&0&0&|&0&0&0&0\\
0&a_1+a_2&0&0&|&-a_2&0&0&0&|&0&0&0&-a_7&|&0&0&a_7 &0\\
0&0&a_5&0&|&0&0&0&0&|&a_9&0&0&0&|&0&0&0&0\\
0&0&0&a_5&|&0&0&0&0&|&0&0&0&0&|&a_9&0&0&0\\
-&-&-&-&-&-&-&-&-&-&-&-&-&-&-&-&-&-&-\\
0&-a_2&0&0&|& a_1+a_2&0&0&0&|&0&0&0&a_7&|&0&0& -a_7 &0\\
0&0&0&0&|&0&a_1&0&0&|&0&0&0&0&|&0&0&0&0\\
0&0&0&0&|&0&0&a_5&0&|&0&a_9&0&0&|&0&0&0&0\\
0&0&0&0&|&0&0&0&a_5&|&0&0&0&0&|&0&a_9&0&0\\
-&-&-&-&-&-&-&-&-&-&-&-&-&-&-&-&-&-&-\\
0&0&a_{10}&0&|&0&0&0&0&|&a_6&0&0&0&|&0&0&0&0\\
0&0&0&0&|&0&0&a_{10}&0&|&0&a_6&0&0&|&0&0&0&0\\
0&0&0&0&|&0&0&0&0&|&0&0&a_3&0&|&0&0&0&0\\
0&-a_8&0&0&|& a_8&0&0&0&|&0&0&0&a_3+a_4&|&0&0&-a_4&0\\
-&-&-&-&-&-&-&-&-&-&-&-&-&-&-&-&-&-&-\\
0&0&0&a_{10}&|&0&0&0&0&|&0&0&0&0&|&a_6&0&0&0\\
0&0&0&0&|&0&0&0&a_{10}&|&0&0&0&0&|&0&a_6&0&0\\
0&a_8&0&0&|&-a_8&0&0&0&|&0&0&0&-a_4&|&0&0&a_3+a_4&0\\
0&0&0&0&|&0&0&0&0&|&0&0&0&0&|&0&0&0&a_3\\
\end{array} \right) \nonumber
\end{eqnarray}
 }

\noindent In section 5 the coefficients ${\rm A},\ldots,{\rm L}$
were introduced to make a comparison with the near plane-wave
S-matrix ${\mathbb S}_{12}^{\rm KMRZ}$ computed in \cite{KMRZ}.
These coefficients are expressed via the coefficients $a_i$ above
as follows \bea
\begin{array}{lllll} {\rm A}=a_1+a_2\, , ~&~
{\rm B}=-a_2\, , ~&~ {\rm C}=-a_8\, , ~&~ {\rm D}=a_3+a_4\, , ~&~ {\rm E}=-a_4 \, ,\\
{\rm F}=-a_7\, ,~&~ {\rm G}=a_{5}\, , ~&~ {\rm H}=a_{10} \, , ~&~
{\rm K}=a_{9}\, , ~&~ {\rm L}=a_6 \, .
\end{array}
\eea Finally, we note that the permutation matrix $P$ corresponds
to the choice
$$
a_1=a_3=a_9=a_{10}=1\, , ~~~~~a_2=a_4 =-1\, ,
~~~~~a_5=a_6=a_7=a_8=0\, ,
$$
while the graded permutation ${\cal P}$ is given by
$$
a_1=a_4=a_9=a_{10}=1\, , ~~~~~a_2=a_3=-1\, ,
~~~~~a_5=a_6=a_7=a_8=0\, .
$$

%%%%%%%%%%%%%%%%%%%%%%%%%%%%%%%%%%%%%%%%%%%%%%%%%%%%%%%%%
\end{document}